\documentclass[fleqn,usenatbib]{mnras}

\usepackage{scalerel,stackengine}
\stackMath
\newcommand\reallywidehat[1]{%
\savestack{\tmpbox}{\stretchto{%
  \scaleto{%
    \scalerel*[\widthof{\ensuremath{#1}}]{\kern-.6pt\bigwedge\kern-.6pt}%
    {\rule[-\textheight/2]{1ex}{\textheight}}
  }{\textheight}%
}{0.5ex}}%
\stackon[1pt]{#1}{\tmpbox}%
}

\usepackage{newtxtext,newtxmath}

\usepackage[T1]{fontenc}
\usepackage{ae,aecompl}
\usepackage{graphicx}

\newcommand{\Ng}{N_{\rm g}}

\newcommand\vr{\vec{r}}

\newcommand\hk{\hat{k}}

\newcommand\hr{\hat{r}}
\newcommand{\vk}{\vec{k}}
\newcommand{\vqo}{\vec{q}_1}
\newcommand{\vqt}{\vec{q}_2}
\newcommand\hqo{\hat{q}_1}
\newcommand\hqt{\hat{q}_2}

\usepackage{amsmath}	
\usepackage{amssymb}	

\title[Decoupling Integrals]{On Decoupling the Integrals of Cosmological Perturbation Theory}

\author[Z. Slepian]{
Zachary Slepian$^{1,2,3}$\thanks{E-mail: zslepian@lbl.gov}
\\
$^{1}$Department of Astronomy, University of Florida, 211 Bryant Space Sciences Center, Gainesville, FL 32611-2055, USA\\
$^{2}$Lawrence Berkeley National Laboratory, 1 Cyclotron Road, Berkeley, CA 94720, USA\\
$^{3}$Berkeley Center for Cosmological Physics, University of California, Berkeley, Berkeley, CA 94720, USA}

\date{Accepted XXX. Received YYY; in original form ZZZ}

\pubyear{2018}

\begin{document}
\label{firstpage}
\pagerange{\pageref{firstpage}--\pageref{lastpage}}
\maketitle

\begin{abstract}
Perturbation theory (PT) is often used to model statistical observables capturing the translation and rotation-invariant information in cosmological density fields. PT produces higher-order corrections by integration over linear statistics of the density fields weighted by kernels resulting from recursive solution of the fluid equations. These integrals quickly become high-dimensional and naively require increasing computational resources the higher the order of the corrections. Here we show how to decouple the integrands that often produce this issue, enabling PT corrections to be computed as a sum of products of independent 1-D integrals. Our approach is related to a commonly used method for calculating multi-loop Feynman integrals in Quantum Field Theory, the Gegenbauer Polynomial $x$-Space Technique (GPxT). We explicitly reduce the three terms entering the 2-loop power spectrum, formally requiring 9-D integrations, to sums over successive 1-D radial integrals. These 1-D integrals can further be performed as convolutions, rendering the scaling of this method $N_{\rm g} \log N_{\rm g}$ with $N_{\rm g}$ the number of grid points used for each Fast Fourier Transform. This method should be highly enabling for upcoming large-scale structure redshift surveys where model predictions at an enormous number of cosmological parameter combinations will be required by Monte Carlo Markov Chain searches for the best-fit values.
\end{abstract}
\begin{keywords}
Cosmology: large-scale structure -- theory
\end{keywords}

\section{Introduction}
In the consensus picture of cosmological structure formation, perturbations that were Gaussian-distributed in amplitude and uniform-randomly-distributed in phase (when written in Fourier space) were generated at the end of inflation (\citealt{Starobinsky:1982}, \citealt{Bardeen:1983}). Baryon Acoustic Oscillations (BAO) in the ionized plasma present prior to decoupling\footnote{When an electron was last scattered by a photon; slightly later than the last scattering of a photon by an electron, which occurred at recombination $z$$\sim$$1100$.} $(z$$\sim$$1020)$ (\citealt{Sakharov:1966}, \citealt{Silk:1968}, \citealt{Sunyaev:1970}, \citealt{Peebles:1970}, \citealt{Bond:1983}, \citealt{Holtzman:1989}, \citealt{Hu:1996}, \citealt{Eisenstein:1998}, \citealt{Weinberg:2002}, \citealt{Eisenstein:2005}, \citealt{Cole:2005}, \citealt{Eisenstein:2007}, \citealt{Slepian_TF:2016}) then imprinted additional spatial correlations on the perturbations, which were then amplified by gravitational instability (e.g. \citealt{Goroff:1986}, \citealt{Jain:1994}) into the large-scale structure observed in low-redshift galaxy surveys (\citealt{Strauss:1995}, \citealt{Coil:2013}, for reviews). It is typically taken that the dark matter density field was dominantly shaped by gravitational evolution, that halos form in this matter field according to their local environment, characterized by bias models (e.g. \citealt{Kaiser:1984}, \citealt{Bardeen:1986}, \citealt{Fry:1993}, \citealt{Mo:1996};  \citealt{Desjacques:2018} for a recent review) or Effective Field Theory (EFT) parameters to be measured from simulations (e.g. \citealt{McDonald:2009}, \citealt{Carrasco:2012cv}, \citealt{Senatore:2015}), and that galaxies form in the dark matter halos according to both environment and detailed gas and astrophysical processes. For purposes of large-scale structure these latter effects are also often compressed into the bias coefficients or EFT parameters.

The evolution of the dark matter field under gravity is a numerically straightforward problem in the sense that the gravitational force law on the relevant scales is well-understood. Nonetheless it is computationally demanding to evolve the matter field in large cosmological volumes for direct comparison with observational proxies for the true density field, given that we do not know the initial 3-D distribution of matter in the Universe (though recent work has made some progress in this direction, e.g. \citealt{Kitaura:2008}, \citealt{Jasche:2010}, \citealt{Jasche:2013}, \citealt{Ata:2015}, \citealt{Leclercq:2015}, \citealt{Seljak:2017}, \citealt{Schmittfull:2017}, \citealt{Jasche:2018}, \citealt{Schmidt:2018}). Thus, as an alternative, translation and rotation-invariant clustering statistics such as the 2-Point Correlation Function (2PCF), power spectrum, 3-Point Correlation Function, and bispectrum, which measure excess clustering over random of galaxy pairs or triplets in respectively configuration and Fourier space, are used (\citealt{Bernardeau:2002}, for a review). These statistics can be predicted by averaging many sub-regions of numerical simulations (though \citealt{Pontzen:2016} and \citealt{Angulo:2016} suggest this can be improved by phase-matching in the initial conditions), but can also be, less accurately but far more quickly, obtained from analytic solution of the approximate equations of motion for the dark matter. These equations of motion are approximate for a number of reasons: they neglect possible velocity dispersion in the CDM trajectories, late-time couplings to other fluids (e.g. neutrinos), and they do not extend beyond shell-crossing, where CDM trajectories cross each other.

The fluid equations are solved in Eulerian coordinates by recursion relations (\citealt{Goroff:1986}, \citealt{Jain:1994}) giving kernels that systematically generate higher-order density corrections as integrals of lower-order fields against the kernels.\footnote{Recursive solutions also exist for Lagrangian PT, although we do not focus on them here; see e.g. \cite{Zhelig:2014} for Einstein-de Sitter, \cite{Rampf_plus:2015} and \cite{Matsubara:2015} for $\Lambda$CDM.} The higher-order corrections to the density can then be used to compute corrections to the clustering statistics. In particular, the corrections to the clustering statistics end up as integrals of products of the kernels against linear power spectra. The advantage is that in the initial field, modulo a small amount of possible primordial non-Gaussianity (PNG), the linear power spectrum contains all of the information. Further, the linear power spectrum can be quickly obtained numerically from linear Boltzmann solver codes such as CMBFAST \citep{Seljak:1996}, CAMB \citep{Lewis:2000}, or CLASS \citep{Lesgourgues:2011}. However, since the linear theory power spectrum is not known in closed form (save for approximately), the relevant integrals must be done numerically, and redone every time a new set of cosmological parameters is desired as these produce a new (numerical) linear power spectrum.

These integrals quickly become high-dimensional, making it numerically cumbersome to compute the corrections. This has not been an entirely limiting factor in previous analyses as often the cosmology was not varied, varied in a way that could be adjusted after this computation (e.g. BAO analyses varying the ratio $\alpha$ of the model to fiducial sound horizon) or varied over only a few fiducial models requiring recomputation. However, for future analyses of upcoming datasets such as DESI \citep{DESI:2016} and LSST \citep{Euclid:2011}, it will be desirable to vary the cosmological parameters an enormous number of times, much as is already done with Monte Carlo Markov Chain (MCMC) analyses of the Cosmic Microwave Background (CMB; e.g. \citealt{Lewis:2002}, \citealt{Planck:2018}). Thus greater speed in computation of the corrections to clustering statistics would be useful.

Several recent approaches to this problem have been proposed. Appendix A of \cite{Slepian_RV:2015} shows that using decoupling into separated 3-D integrals and then factorization into radial and angular pieces, certain fourth order terms in the 2PCF (1-loop corrections) can be computed as 1-D Hankel transforms by performing the angular integrations analytically.\footnote{A Hankel transform is just an integral of a given function against a spherical Bessel function with a particular power-law weight.} \cite{Ferraro:2012} also exploited this idea even earlier (personal communication) to obtain their equations 13-15 although the details of the derivation do not appear in the paper. Angular-radial factorizations also substantially simplify the tree-level 3PCF into products of 1-D transforms of the power spectrum (e.g. \citealt{Slepian_RV:2015} equations 51-56), as might be expected since the tree-level 3PCF is also fourth order and, like the 1-loop 2PCF, just involves 1-loop kernels (some of these transforms become 2-D in redshift space; \citealt{Slepian_RSD_model:2016} \S2.2). 

 \cite{Schmittfull_1loop:2016} develops this idea much more fully and demonstrates all 1-loop corrections to the power spectrum  can be computed in this way. \cite{Schmittfull_2loop:2016} extends this analysis to 2-loop corrections, but finds that several of the 2-loop terms are not amenable to the methods of separation employed in the previous works above and thus introduces a dummy free variable to be integrated over at the end subject to a Dirac Delta function. This trick allows formal separation. Thus at 2-loop one requires 2-D integrals for a number of terms; indeed one has an infinite sum over such integrals (though it converges quickly in practice). These 2-D integrals dominate the computational cost of the full 2-loop calculation (\citealt{Schmittfull_2loop:2016} \S V.D). 

\cite{McEwen:2016} and \cite{Fang:2017} combine these ideas with expressing the power spectrum as a sum of complex power laws to further accelerate the 1-loop power spectrum computations; the second of the above extends this to non-isotropic quantities relevant for working in redshift space. In this latter case, one introduces velocity as well as density kernels, but they are the same in form with different coefficients so the same techniques apply. Another complication in redshift space is having a preferred direction (the line of sight) and projecting onto it and transverse to it. So doing simply introduces additional angular dependences to the integrals, but these are known in closed form and can be factorized and dealt with analytically. 

\cite{Simonovic:2018} takes a related approach, showing that the integrals required for power laws can all be done analytically, meaning that the challenging part, that of the coupled-multi-dimensional integrals, need only be done once, and then assembled in the linear combination dictated by the power law weights of the initial decomposition of the linear power spectrum. \cite{Assassi:2017} uses this approach for angular statistics that come from integrating the power spectrum against spherical Bessel functions (sBFs), as does \cite{Gebhardt:2018}. Both of these works express the double spherical Bessel function integrals involved in the angular statistics in terms of hypergeometric functions, and the latter particularly focuses on developing a stable, fast, and accurate recursive approach to their computation.

\cite{Fonseca:2017} uses factorization and decoupling to reduce the 1-loop power spectrum computation to 1-D integrals in a slightly different way than \cite{Schmittfull_1loop:2016} and \cite{McEwen:2016}. Writing the Fourier space integral for the loop correction with a Dirac Delta function to enforce momentum conservation rather than as an explicit convolution, they then rewrite the Delta function as the inverse FT of unity. This trick is also used in \cite{Schmittfull_2loop:2016} for the 2-loop integrals. The inverse FT's plane waves can be expanded into sBFs and spherical harmonics (their \S3.2), and the integrals over sBFs done analytically (their Appendix B).\footnote{Their Appendix is also a rather complete reference for previous work on analytic evaluation of multiple sBF integrals, although attention should also be drawn to a method for evaluating triple-sBF integrals by recursion in the Appendix of \cite{Wang:2000}, which they do not mention. \cite{Adkins:2013} also has a number of interesting results on singular integrals of sBFs.} This resolution of the Delta function was employed earlier in \cite{Slepian_3pt_alg:2015} (their equation 58) to reduce the disconnected piece of the 3PCF covariance to tractable 2-D integrals. As already mentioned earlier, \cite{Schmittfull_2loop:2016} also used a Delta function for the 2-loop calculation, although with a different expansion of it than \cite{Fonseca:2017} use for the 1-loop calculation. Rewriting the Dirac delta function is also used in \cite{Bohm:2016} (Appendix B footnote 6). This latter work also employs angular-radial decoupling (equations B1 and B2) and products in real space to reduce 3-D convolutions to integrals of 1-D product integrals (equation B10). Factorization of the Delta function by rewriting as an inverse FT is also used to accelerate the bispectrum algorithm of \cite{Scoccimarro:2015}; we will further discuss algorithms taking advantage of factorization slightly later. \cite{Schmidt:2018} shows that all of the perturbation theory corrections and biasing terms at 1-loop can be written in terms of a set of 28 independent integrals; their Appendix H discusses the fast evaluation of these integrals exploiting the same approach of using configuration space to perform products.\footnote{Another, less-well-known but useful case where a 3-D computation in configuration space can be simplified with analytic angular integration is \cite{Zehavi:2005} equation 5 for $\sigma_R$, the standard deviation of the overdensity smoothed to a given scale $R$. $\sigma_R$ can be written as a 1-D integral in configuration space, despite that it naively appears to require a convolution there. The trick is to convolve geometrically as the overlap lens between two spheres. Of course $\sigma_R$ is standardly expressed as a 1-D integral in Fourier space, but working in configuration space is useful if one wishes to compute the variance of data. One then avoids needing to grid the data for an FT. A similar geometric trick is used in \cite{Hand:2017} \S3.3. to examine a toy-model of boundary effects on the power spectrum multipoles.}

\cite{Taruya:2018} exploits the fact that, given a realization of the initial, Gaussian, linear density and velocity fields, computation of higher-order corrections to the fields themselves (as opposed to their statistics) is a series of 3-D forward and inverse FTs. This work thus achieves $\Ng \log \Ng$ evaluation of these corrections, where $\Ng$ is the number of grid points used for the FT. However the price of this method is cosmic variance: many realizations must be averaged over to obtain precise density field statistics. McDonald (personal communication) also shows that the 1-loop density and velocity field corrections and their statistics can be expressed with configuration space products rather than the Fourier-space convolutions the PT recursion relations imply. The method presented in that work is analogous to the split-step Fourier method in spectral solution of partial differential equations, where one takes diffusion steps in Fourier space but time-steps in configuration space (e.g. \citealt{Agrawal}). McDonald suggests taking products in configuration space but applying derivative operators in Fourier space. 

The purpose of the present work is evaluating integrals in cosmological perturbation theory; of course, this is only worthwhile if PT offers a reasonable description of the clustering of matter. Full discussion of the convergence of PT is beyond the scope of this work. In 3-D, the corrections rapidly become complicated, with a large number of terms. \cite{Zhelig:2014} (Einstein-de Sitter) and \cite{Rampf_plus:2015} ($\Lambda$CDM) show the convergence in 3-D up to a certain time away from the initialization. In 1-D corrections of arbitrarily higher order are more tractable, rendering it easier to compare the fully-summed PT calculation to simulations. Several works have investigated convergence from this perspective: \citealt{Novikov:1969}, \citealt{McQuinn:2016}, \citealt{Taruya:2017}, \citealt{Pietroni:2018}. We also point to \citealt{Rampf:2017} which works in quasi-1-D. \cite{McQuinn:2016} finds that it converges but not to the correct answer, although \cite{Pietroni:2018} suggest adding non-perturbative terms is able to greatly improve the agreement, a point also implicit in the discussion of \cite{Rampf:2017}. \cite{Pajer:2018} extend the work of \cite{McQuinn:2016} to non-Gaussian initial conditions and the bispectrum, as well as showing that divergences occur in the configuration-space 2PCF. \cite{Rampf_pert:2017} analyses the quasi-spherical collapse of a perturbed tophat, finding that generically collapse occurs earlier and the linear density at collapse is reduced relative to the spherical case. \citealt{Saga:2018} investigates Lagrangian PT in a semi-3-D limit that might be described as $1+1+1$, i.e. halos are modeled as being seeded by three crossed sine waves along the three Cartesian axes. This model allows derivation of recursion relations that are simple enough to evaluate to high (tenth) order and can be compared with simulations; reasonable agreement is found. 

\cite{McDonald:2018} presents a new method in principle applicable in 3-D although only tested numerically in 1-D thus far. The method performs well in 1-D, going beyond the point where PT traditionally breaks down (shell-crossing, where the determinant of the Jacobian between Eulerian and Lagrangian space becomes singular). It is inspired by the Hubbard-Stratonovich transformation typically used in Quantum Field Theory (QFT), which is in \cite{McDonald:2018} used to carry an exact exponential of the displacement field farther into the calculation, allowing increased accuracy. Motivated by this work we investigated use of a Hubbard-Stratonovich transformation to simplify the integrals of perturbation theory treated here, but without success. Overall, application of techniques from QFT to cosmological perturbation theory is a rich subject of future investigation.\footnote{Suggestively, \cite{Simonovic:2018} notes that the 2-loop standard PT integrals are equivalent to those in a massless scalar QFT with a cubic interaction.} The method presented in the present work is in fact similar, up to a point, to a well-known method for computing multi-loop Feynman integrals called the Gegenbauer Polynomial $x$-Space Technique (GPXT; \citealt{Chetyrkin:1980}). We return to this and other connections to QFT in \S\ref{subsec:QFT}.

The idea of factorization has also been more generally exploited to develop fast algorithms for measuring the clustering of large-scale structure. \cite{Slepian_3pt_alg:2015, Slepian_FT_3PCF:2016, Slepian_aniso:2017}\footnote{Implementations of these latter two are presented in \cite{Portillo:2018} and \cite{Friesen:2017}.}, and \cite{Sugiyama:2018} all use splitting of Legendre polynomials into spherical harmonics by the addition theorem (\ref{eqn:sph_add_thm}) to accelerate measurement of the 3PCF as well as the anisotropic 2PCF, and \cite{Hand:2017} also exploits this factorization to develop a fast algorithm for measuring the anisotropic power spectrum to high multipole.\footnote{Publicly available through \textsc{nbodykit}; see \citealt{Hand:2018}.} \cite{Bianchi:2015} also uses an angular-radial splitting to measure the anisotropic power spectrum, although their basis is not orthogonal and so ends up requiring more FTs than \cite{Hand:2017}.

In this work we take a somewhat different tack from the PT works discussed above although conceptually the motives are the same. Much as these earlier works largely do, we seek to exploit the isotropy of the power spectrum in combination with angular-radial factorization to perform as many integrals as possible analytically just once. We also require a general way of decomposing the types of integration kernel that generically lead to the coupled nature of the loop integrals, and thus their high dimensionality and numerical costliness. 

All of the 2-loop contributions are over two linear momenta. If certain coupled terms could be decoupled, each of these two momentum integrals can be written as a convolution, one nested inside the other. This idea is similar to that in \cite{Taruya:2018} except at the level of the density field statistics rather than the raw field. The nested double convolution gives one four different arguments in the integrand ``for free,'' two for each convolution, for instance $\vec{q}_1, \vec{q}_1 + \vec{a}$ and $\vec{q}_2, \vec{q}_2 + \vec{b}$ if one is integrating over dummy momenta $\vec{q}_1$ and $\vec{q}_2$. The convolutions can then be factored into radial and angular pieces and reduced to 1-D radial integrals against the linear power spectrum. 

Our main problem is to decouple the coupled terms in the 2-loop integrands that, left unaddressed, spoil the double convolutionality. We need both to decouple them and then factorize them into angular and radial pieces to be able to incorporate them into the convolutional terms in the integrand. There are six arguments in the 2-loop integrals, and as noted above, convolutions deal with four ``for free''; hence we will need to decouple at most two terms.  We note that throughout this work, we use the word ``decouple'' to denote pulling apart two 3-D vectors so they may be integrated over separately, and the word ``factorize'' to denote separating radial and angular dependences within functions of a single 3-D vector. We also use the word decouple to indicate removing a constraint on the magnitude of one 3-D vector relative to that of another, i.e. that it be less than the other; our meaning should be clear from context.

This work is laid out as follows. In \S\ref{sec:casting} we present the general integrals we treat and show that if certain parts of the integrand could be decoupled the rest of the calculation becomes a set of nested 3-D convolutions. In \S\ref{sec:decoupling} we show how to decouple these parts of the integrands by deriving a decoupled and factorized eigenfunction expansion for them. In \S\ref{sec:as_convols} we insert these expansions and show the whole calculation is now a set of 3-D convolutions. In \S\ref{sec:evals} we evaluate these convolutions, ultimately reducing them to an infinite sum over 1-D integrals that can be done quickly using FFTs. \S\ref{sec:disc} discusses possible extensions of our results, connections to QFT, and outlines how a numerical implementation might proceed. We conclude in \S\ref{sec:concs}.

\section{Casting the 2-Loop Integrals as Double Convolutions}
\label{sec:casting}
Throughout, our convention will be that forward 3-D FTs will have a positive $i$ in the plane wave, and inverse FTs a negative $i$ and be normalized by $(2\pi)^{-3}$.  
\cite{Schmittfull_2loop:2016} express the three 2-loop integrals as 
\begin{align}
&\mathcal{I}_{ij, {\rm SV}}(k,\vec{\alpha},\vec{\beta}) =\nonumber\\
&\int \frac{d\Omega_k}{4\pi} \int \frac{d^3\vec{q}_1}{(2\pi)^3} \frac{d^3\vec{q}_2}{(2\pi)^3} \;\frac{ e^{i\vec{\alpha}\cdot \vec{q}_1}e^{i\vec{\beta}\cdot \vec{q}_2}  }{q_1^{2n_1} |\vec{k} + \vec{q}_1|^{2n_1'}  q_2^{2n_2} |\vec{k}+\vec{q}_2|^{2n_2'}  }  \nonumber\\
&\times \frac{P_{\rm lin}(q_1)P_{\rm lin}(q_2)}{ |\vec{q}_1 + \vec{q}_2|^{2n_3}  |\vec{k} + \vec{q}_1 + \vec{q}_2|^{2n_3'}} P_{\rm lin}(|\vec{w}_{ij}|)
\label{eqn:I15_SV}
\end{align}
with
\begin{align}
\vec{w}_{15} = \vec{k},\;\;\vec{w}_{24} = \vec{k} + \vec{q}_2,\;\;\vec{w}_{33} = \vec{k} + \vec{q}_1 + \vec{q}_2.
\label{eqn:w_ij}
\end{align}
Relative to \cite{Schmittfull_2loop:2016} we have added averaging over $d\Omega_k$, which they presumably intended since the power spectrum and its loop corrections are isotropic (ignoring for the moment redshift-space distortions (RSD)). The vectors $\vec{\alpha}$ and $\vec{\beta}$ are parameters with respect to which one can differentiate to generate more complicated numerators, and then take the limit $\vec{\alpha}, \vec{\beta} \to 0$. We note that any dependence on $\vec{q}_1$ and $\vec{q}_2$ in the numerator must be rotation-invariant as the PT kernels cannot depend on absolute direction. We also pause to note that these integrals are not divergent: for instance, by momentum conservation, if $\vec{q}_1 = -\vec{q}_2$, then $k \to 0$ and $P_{\rm lin} (0) = 0$. An analogous point applies to the other denominators that could potentially vanish.

We notice that all three 2-loop integrals have largely the same structure, with just a different argument of the third linear power spectrum. We choose to drop the exponentials in favor of writing out a Legendre series for the terms in the numerator they are meant to capture. We have
\begin{align}
&N(\vec{q}_1, \vec{q}_2) = \sum_{\ell=0}^{\ell_{\rm max}} N_{\ell}(q_1,q_2)\mathcal{L}_{\ell}(\hat{q}_1\cdot\hat{q}_2)\nonumber\\
&=\sum_{\ell=0}^{\ell_{\rm max}} N^{[1]}_{\ell}(q_1) N^{[2]}_{\ell}(q_2)\mathcal{L}_{\ell}(\hat{q}_1\cdot\hat{q}_2)\nonumber\\
&=\sum_{\ell=0}^{\ell_{\rm max}} N^{[1]}_{\ell}(q_1) N^{[2]}_{\ell}(q_2)\frac{4\pi}{(2\ell + 1)}\sum_{m = -\ell}^{\ell}Y_{\ell m}(\hat{q}_1) Y^*_{\ell m}(\hat{q}_1).
\label{eqn:leg_numer}
\end{align}
Noticing that any parametric differentiation of the exponentials in equation (\ref{eqn:I15_SV}) would give rise to a factorizable function of $q_1$ and $q_2$ allows us to assert the second equality above, and in the third line we have factored the angular dependence using the spherical harmonic addition theorem (\ref{eqn:sph_add_thm}). We do note that the numerator may need an additional sum at each $\ell$ to be factored radially; a simpler example of this type of behavior is in the dipole term of the $F^{(2)}$ kernel, which is $(1/2) (q_1/q_2 + q_2/q_1) (\hat{q}_1 \cdot \hat{q}_2) $. This extra layer will not alter any of our results as it can be incorporated as a trivial sum at the end of our calculations. 

The fundamental set of integrals we will pursue here thus becomes
\begin{align}
&\mathcal{I}_{ij}(k) = \sum_{\ell=0}^{\ell_{\rm max}} \sum_{m = -\ell}^{\ell}\frac{4\pi}{2\ell + 1}\int \frac{d\Omega_k}{4\pi} \int \frac{d^3\vec{q}_1}{(2\pi)^3} \frac{d^3\vec{q}_2}{(2\pi)^3} \;\nonumber\\
&\times \frac{ N_{\ell}^{[1]}(q_1)  N_{\ell}^{[2]}(q_2) Y_{\ell m}(\hat{q}_1) Y^*_{\ell m}(\hat{q}_1)}{q_1^{2n_1} |\vec{k} + \vec{q}_1|^{2n_1'}  q_2^{2n_2} |\vec{k}+\vec{q}_2|^{2n_2'}  }  \nonumber\\
&\times \frac{P_{\rm lin}(q_1)P_{\rm lin}(q_2)}{ |\vec{q}_1 + \vec{q}_2|^{2n_3}  |\vec{k} + \vec{q}_1 + \vec{q}_2|^{2n_3'}} P_{\rm lin}(|\vec{w}_{ij}|)
\label{eqn:Ifun}
\end{align}
and for compactness of notation we define all terms but the outer sums and the $\ell$-dependent pre-factor as $\mathcal{I}_{ij, \ell m} (k)$ so that
\begin{align}
&\mathcal{I}_{ij}(k) = \sum_{\ell=0}^{\ell_{\rm max}}\sum_{m = -\ell}^{\ell}\frac{4\pi}{2\ell + 1} \mathcal{I}_{ij, \ell m}(k).
\end{align}
We will focus on the $\mathcal{I}_{ij, \ell m}$ as then summing over $\ell$ and $m$ at the end is a trivial additional step. We now introduce a kernel $K$ such that
\begin{align}
&\mathcal{I}_{ij, \ell m}(k) = \nonumber\\
&\int \frac{d\Omega_k}{4\pi} \int \frac{d^3\vec{q}_1}{(2\pi)^3} \frac{d^3\vec{q}_2}{(2\pi)^3}\; 
K(\vec{k}, \vec{q}_1, \vec{q}_2)  P_{\rm lin}(|\vec{w}_{ij}|).
\end{align}
In other words, we may take advantage of the fact that the three integrals (15, 24, and 33) differ only with respect to the argument of the final power spectrum to focus on integrals of that final power spectrum against the same kernel $K$ for all three.

We seek to cast each of the $\mathcal{I}_{ij, \ell m}$ as double 3-D convolutions, with the inner integral over $\vec{q}_2$ a 3-D convolution the result of which is then further convolved with additional factors as we perform the outer integral over $\vec{q}_1$. In short, we wish to render the computation ``doubly convolutional.''

Suppose we could find a way to split $K$, perhaps differently for each of the three 2-loop terms, into a piece that would be explicitly doubly convolutional and another piece that, if we could factor it into separated functions of $\vec{k}, \vec{q}_1,$ and $\vec{q}_2$, could also be incorporated into the double convolutions. Pursuing such a splitting, we write
\begin{align}
K = K_{ij}^{\rm c}K_{ij}^{\rm d}
\end{align}
where ${\rm c}$ is for ``convolutional'' and ${\rm d}$ is for ``decoupled.'' We have added subscripts to the righthand side to indicate that the required splitting may be different for each of the three 2-loop contributions.

For $\mathcal{I}_{15}$, we set 
\begin{align}
K_{15}^{\rm c} &= \frac{ N_{\ell}^{[1]}(q_1) Y_{\ell m}(\hat{q}_1) P_{\rm lin}(q_1) N_{\ell}^{[2]}(q_2) Y^*_{\ell m}(\hat{q}_2) P_{\rm lin}(q_2)}{q_1^{2n_1} |\vec{k} + \vec{q}_1|^{2n_1'}  q_2^{2n_2} |(\vec{k} + \vec{q}_1) + \vec{q}_2|^{2n_3'} }\nonumber\\
&= \mathcal{R}_{\ell}^{n_1}(q_1) \mathcal{P}^{n_1'}(\vk + \vqo)\mathcal{R}_{\ell }^{n_2}(q_2)\mathcal{P}^{n_1'}((\vk + \vqo) + \vqt)
\label{eqn:Kc_15}
\end{align}
where we have defined
\begin{align}
&\mathcal{R}_{\ell}^{n}(q_t) = N_{\ell}^{[t]} (q_t)q_t^{-2n} P_{\rm lin}(q_t),\nonumber\\ 
& \mathcal{P}^{n}(\vec{q}) = | \vec{q}|^{-2n},
\label{eqn:calRP}
\end{align}
with $\mathcal{R}$ for ``radial'' and $\mathcal{P}$ for ``power law''.  We have included a subscript in the argument of $\mathcal{R}$ becauase there it should match the superscript of $N$.
We set
\begin{align}
K_{15}^{\rm d} = \frac{1}{|\vec{q}_1 + \vec{q}_2|^{2n_3}  |\vec{k}+\vec{q}_2|^{2n_2'}}. 
\label{eqn:Kd_15}
\end{align}
We have rearranged the factors in equation (\ref{eqn:Kc_15}) relative to those in equation (\ref{eqn:Ifun}) so as to group the terms in $\vqo$ together and the same for those in $\vqt.$ $K_{15}^{\rm c}$ is clearly convolutional in that we could integrate over $\vec{q}_2$ at an offset of $\vec{k} + \vec{q}_1$. We could then integrate that result against $\vec{q}_1$, leading to a convolution over $\vec{q}_1$ at an offset $\vec{k}$. We note that the third power spectrum in equation (\ref{eqn:Ifun}), which for $\mathcal{I}_{15}$ becomes $P_{\rm lin}(k)$, can be pulled outside these integrals.

For $\mathcal{I}_{24}$ we are motivated by the argument of the third power spectrum (i.e. $\vec{w}_{24}$ in equation \ref{eqn:w_ij}) to define 
 \begin{align}
&K_{24}^{\rm c} = \mathcal{R}_{\ell }^{n_1}(q_1) \mathcal{P}^{n_1'}(\vk + \vqo)\mathcal{R}_{\ell }^{n_2}(q_2)\mathcal{P}^{n_1'}(\vk  + \vqt)
\end{align}
and 
\begin{align}
K_{24}^{\rm d} = \frac{1} { |\vqo + \vqt|^{2n_3} |  \vqo + (\vk  + \vqt) |^{2n_3'} }. 
\label{eqn:Kd_24}
\end{align}
$K_{24}^{\rm c}$ is convolutional: we may first convolve just the $\vec{q}_2$-dependent factors over $\vqt$ at an offset of $\vec{k}$, obtaining a result dependent only on $\vk$. We note that the power spectrum has argument $\vk + \vqt$ (see equation \ref{eqn:Ifun}) and so could also be included in this inner convolution. We may then perform a separate convolution of the $\vqo$-dependent factors over $\vqo$. We would then finally integrate the product of the inner and outer convolution results over $\vk$.

Finally, motivated by $\vec{w}_{33} = \vec{k} + \vec{q}_1 + \vec{q}_2$ in the third power spectrum in $I_{33}$, we set
\begin{align}
K_{33}^{\rm c}  = K_{15}^{\rm c},\;\; 
K_{33}^{\rm d}  = K_{15}^{\rm d}. 
\label{eqn:Kd_33}
\end{align}
We note that the power spectrum in $\vk  + \vqo + \vqt$ can be included in the inner convolution.

\begin{figure}
\centering
    \includegraphics[width=.5\textwidth]{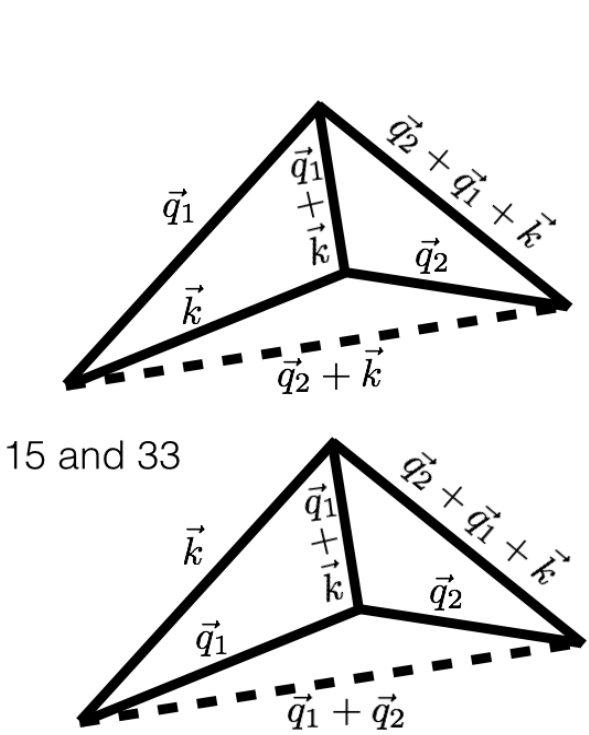}
	\caption{Diagram for 15 and 33 contributions. The solid lines show the pieces of the 15 and 33 integrands that are already convolutional; the two dashed lines show those that need to be decoupled. Both $\mathcal{I}_{15}$ and $\mathcal{I}_{33}$ require both diagrams each, although the convolutional part of the integrands does not change. We simply had to flip two of the vectors to be able to draw the two different non-convolutional, coupled pieces.}
\label{fig:15_33}
\end{figure}

\begin{figure}
	\centering
    \includegraphics[width=.4\textwidth]{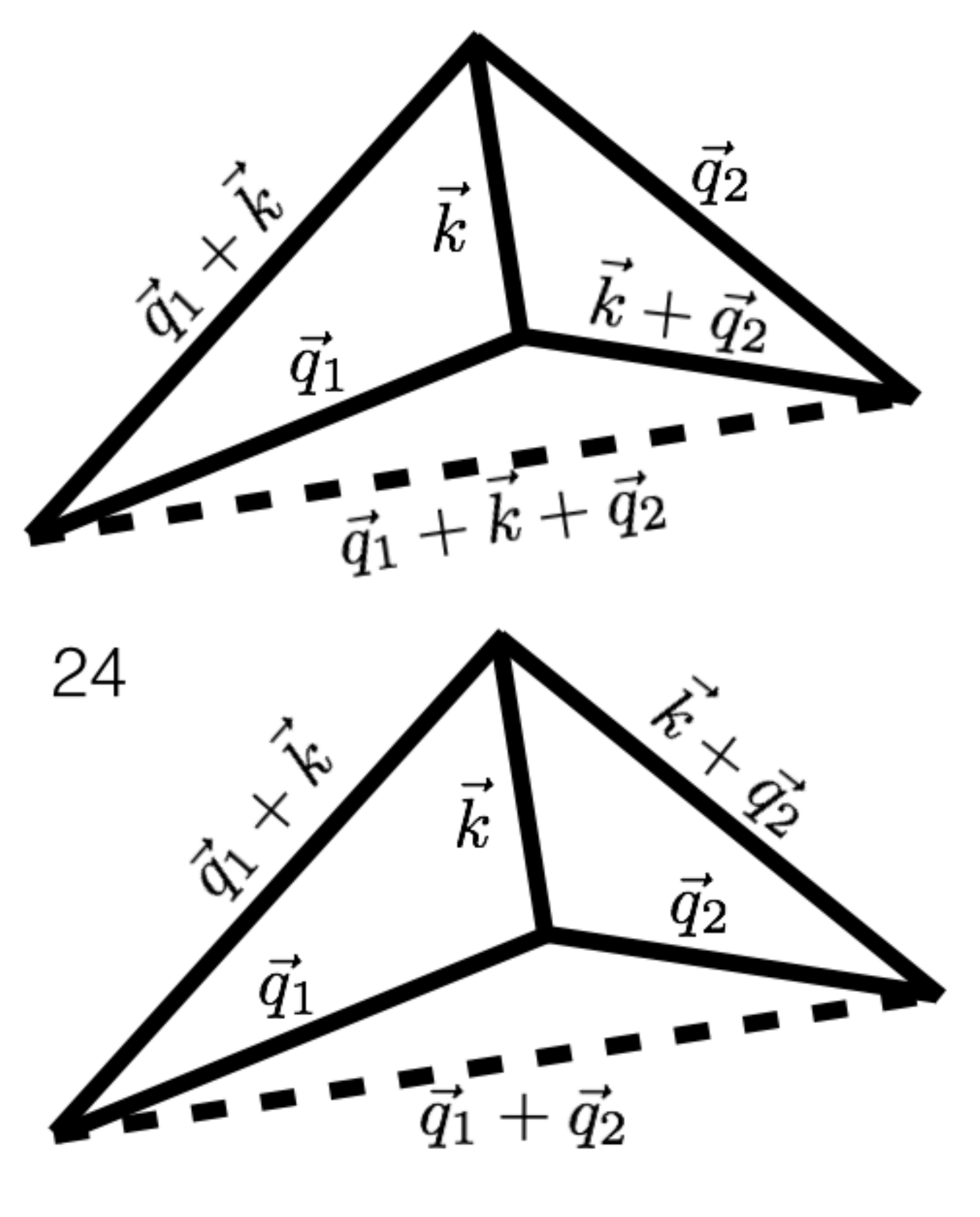}
	\caption{Diagram for 24 contribution. The solid lines show the pieces of the 24 integrand that are already convolutional; the two dashed lines show those that need to be decoupled. $\mathcal{I}_{24}$ requires both diagrams, although the convolutional part of the integrand does not change. We simply had to flip two of the vectors to be able to draw the two different non-convolutional, coupled pieces.}
\label{fig:15_33}
\end{figure}

\section{Decoupling}
\label{sec:decoupling}
We now have two tasks. First, we need to obtain explicit representations of the $K^{\rm d}$ that are separated into functions of one of its three arguments each. Second, we need to factor both the convolutional ($K^{\rm c}$) and decoupled ($K^{\rm d}$) kernels into radial and angular pieces. We may  then exploit the isotropy of the power spectrum, the only component that is not available in closed form, to evaluate the angular dependence analytically. Doing so will reduce the numerical integrals we must do to 1-D radial transforms that can be done efficiently.

We discuss each task in turn, first considering the decoupling. \cite{Schmittfull_2loop:2016} have the powers $n_t$, $t \in \{1, 1', 2, 2', 3, 3'\}$, take on values of $1$ and $2$ (see equation \ref{eqn:I15_SV}), so we need to decouple denominators of the form
\begin{align}
\frac{1} {|\vec{p}_1 + \vec{p}_2|^2},\;\;\frac{1}{|\vec{p}_1 + \vec{p}_2|^4},
\end{align} 
where $\vec{p}_1$ and $\vec{p}_2$ stand in for any of $\vec{k},\; \vec{q}_1$, and $\vec{q}_2$.
 
We take motivation from \cite{Limpanuparb:2011} and \cite{Dominici:2012}. These works use a multipole expansion to write
\begin{align}
\frac{1} {|\vec{p}_1 + \vec{p}_2|} = \sum_{L = 0}^{\infty} \frac{p_<^{L}}{p_>^{L+1}}\mathcal{L}_{L}(\hat{p}_1\cdot \hat{p}_2).
\label{eqn:multipole_series}
\end{align}
However, in equation (\ref{eqn:multipole_series}) the two momentum magnitudes are only ${\it formally}$ decoupled; in reality, one must be constrained by the other when we integrate to insure that the lesser-greater restriction on the right-hand side is met. Thus one cannot do the radial integrals over $p_1$ and $p_2$ in a truly separated way, so the computational cost remains $N_{\rm g}^2$ rather than $2N_{\rm g}$, with $N_{\rm g}$ the number of grid points in the $p_t$, $t = 1, 2$ used for the integration. We require a means of truly decoupling the radial part of the multipole expansion (\ref{eqn:multipole_series}).

\cite{Limpanuparb:2011} and \cite{Dominici:2012} do this via the integral
\begin{align}
&\frac{2}{\pi}(2 L +1)\int dx\; j_{L}(p_1x)j_{L}(p_2x) \nonumber\\
&= \frac{p_1^{L}}{p_2^{L +1}},\;\;p_1< p_2,\;\;\frac{p_2^{L}}{p_1^{L +1}},\;\;p_2< p_2.
\label{eqn:dec_integral_Limp}
\end{align}
We prove this integral in Appendix \ref{app:dec_int_1_proof}. Rewriting the radial piece of the multipole series (\ref{eqn:multipole_series}) via the integral (\ref{eqn:dec_integral_Limp}) thus offers a true rather than merely formal decoupling: but at the price of an additional integral over $x$ at the end. 

\cite{Limpanuparb:2011} and \cite{Dominici:2012} circumvent this issue by finding that the integral is exactly equal to an infinite sum. This is essentially a sampling rule for performing the integral numerically; one can take this identity to mean that taking the integrand at integer values only of $x$ is the best integration scheme. In particular, \cite{Dominici:2012} show the integral-sum identity that
\begin{align}
\int dx\; j_{L}(p_1x)j_{L}(p_2x) = \sum_{n = 0}^{\infty} n\epsilon_n\; j_{L}(np_1) j_{L}(np_1)
\label{eqn:dec_int_1}
\end{align}
where we have taken their Corollary 3.3 and rewritten in terms of spherical Bessel functions using that $j_L(x) = \sqrt{\pi /(2x)}J_{L + 1/2}(x)$. Following \cite{Dominici:2012} we have defined
\begin{align}
\epsilon_n = 1,\;\;n>0;\;\;=1/2,\;\; n = 0.
\end{align}
The identity (\ref{eqn:dec_int_1}) leads them to the decomposition 
\begin{align}
&\frac{1} {|\vec{p}_1 + \vec{p}_2|} = \sum_{L=0}^{\infty} (2L+1)\sum_{n=0}^{\infty} \phi_{nL M} (\vec{p}_1)\phi^*_{nL M}(\vec{p}_2),\nonumber\\
&\phi_{n L M} (\vec{p}) = \sqrt{n \epsilon_n}j_{L}(np) Y_{LM}(\hat{p}),
\label{eqn:Dominici_decomp}
\end{align}
which has the additional advantage of being factorized into radial and angular pieces.  We term the integral (\ref{eqn:dec_integral_Limp}) a ``decoupling integral.''  We note that the integral-to-sum identity is only valid in the range $0< p_1,\; p_2 < \pi$, but since the power spectrum has effectively compact support due to the damping from non-linear structure formation, we can always rescale our integration domain of momentum magnitudes appropriately.

To adopt this approach to decoupling here, we need to solve three problems. First is to find an appropriate series expansion of $1/|\vec{p}_1 + \vec{p}_2|^2$ and $1/|\vec{p}_1 + \vec{p}_2|^4$ to be used in the kernels $K^{\rm d}$, in essence a generalization of the multipole expansion for powers other than the inverse first power. Second is to find a decoupling integral giving the right ratio of powers of the momentum magnitudes for the series developed in the first step. Third is to hope that this decoupling integral has an exact integral-to-sum identity like that \cite{Limpanuparb:2011} and \cite{Dominici:2012} exploit.

These three problems are surmountable and in the following subsection we discuss how. 

 \subsection{Decoupling $K^{\rm d}$}
Motivated by the discussion above, we adopt the generalization of Legendre polynomials, Gegenbauer polynomials $C^{(\lambda)}_{L}$, and use the expansion
\begin{align}
\frac{1}{|\vec{p}_1 + \vec{p}_2|^{2\lambda}} = \sum_{L = 0}^{\infty} \frac{p_<^{L}}{p_>^{L + 2\lambda}}C_{L}^{(\lambda)}(\hat{p}_1 \cdot \hat{p}_2) 
\label{eqn:gen_exp}
\end{align}
(e.g. \citealt{AWH13} or \citealt{Sack:1964} equation 4).  It is clear that this form will cover the required cases for powers of the vector sums we wish to decompose, but we also need to decouple the angular dependence. This decoupling can be done using the addition theorem for Gegenbauer polynomials (e.g. \citealt{Koornwinder:1977} equation 3.1). This theorem is a special case of the addition theorem for Jacobi polynomials $P^{\nu, \mu}_n$, which reduce to Gegenbauer polynomials for $\nu = \mu$. However, the Gegenbauer polynomial addition theorem expresses the Gegenbauer polynomial of a dot product as a sum over products of Gegenbauers of the individual angles as well as powers of sines of the individual angles. 

For our case it will be more convenient to use a ``mixed'' addition theorem we prove in Appendix \ref{app:g_add_thm}, which is to our knowledge novel to this work. We decompose a Gegenbauer polynomial of a dot product $x \equiv \hat{a} \cdot \hat{b}$ into a finite sum of products of spherical harmonics each of only one unit vector. We have
\begin{align}
C_{L}^{(\lambda)}(x) = \sum_{J=0}^{L} w_J^{L, \lambda} \sum_{S = -J}^J  Y_{JS}(\hat{a})Y_{JS}^*(\hat{b}),
\end{align}
where $w_J^{L, \lambda} $ is a constant coefficient defined in equation (\ref{eqn:C_thm}).

We note that a seeming alternative would have been to use parametric differentiation with respect to $\cos \theta_{12} \equiv \hat{p}_1 \cdot \hat{p}_2$ on the Legendre series (\ref{eqn:multipole_series}), having written $|\vec{p}_1 + \vec{p}_1| = \sqrt{ p_1^2 + p_2^2 + 2 p_1 p_2 \cos \theta_{12}}$ in the denominator. However this will only raise the denominator's power by even steps, so we could generate $|\vec{p}_1 + \vec{p}_2|^{-3},|\vec{p}_1 + \vec{p}_2|^{-5}$, etc. but never the even powers (2 and 4) we need.

We now seek the decoupling integral to rewrite the radial part of the expansion (\ref{eqn:gen_exp}), concentrating first on $\lambda = 1$. After some experimentation, we find that the integral of the symmetrized sum
\begin{align}
&\frac{2}{\pi}\int_0^{\infty} dx\;x \left[j_{L + 1}(xp_1) j_{L}(xp_2) +  j_{L + 1}(xp_2) j_{L}(xp_1)\right]\nonumber\\
 &=\frac{p_2^{L}}{p_1^{L + 2}},\;\;p_1>p_2,\;\;\frac{p_1^{L}}{p_2^{L + 2}},\;\;p_2>p_1.
 \label{eqn:second_dec_int}
\end{align}
The second term integrates to zero when $p_1>p_2$, and the first term integrates to zero when $p_2>p_1$, so in combination they always provide the desired ratio. We prove this integral in Appendix \ref{dec_int_2_proof}.

We now need a suitable integral-to-sum identity to convert the decoupling integral (\ref{eqn:second_dec_int}) to a sum. \cite{Dominici:2012} supplies one, though with the restriction that one free argument (e.g. $p_1$) be less than the other (e.g. $p_2$). Of course this is exactly the restriction we sought to relax in moving to a decoupling integral in the first place. However it can be shown that the restriction under which \cite{Dominici:2012} proves the integral can be relaxed to cover the cases we need; details are given in Appendix \ref{app:identity_relax}.

The integral-to-sum identity discussed there, with our extension to  cover both $p_1< p_2$ and $p_2 <p_1$, gives
\begin{align}
&\frac{2}{\pi} \int_0^{\infty} dx\;x \left[j_{L + 1}(xp_1) j_{L}(xp_2) +  j_{L + 1}(xp_2) j_{L}(xp_1)\right]\\
&= \frac{4}{\pi^2}  \sum_{n=0}^{\infty} n  \epsilon_n\; \sqrt{p_1 p_2} \left[j_{L + 1}(np_1) j_{L}(np_2) + j_{L  + 1}(np_2) j_{L}(np_1)\right].\nonumber
\end{align}
Consequently we see that
\begin{align}
\label{eqn:denom_decoup_sq}
&\frac{1}{|\vec{p}_1 + \vec{p}_2|^{2}} =\frac{4}{\pi^2} \sum_{L = 0}^{\infty} \sum_{n=0}^{\infty} n \epsilon_n\; \sqrt{p_1 p_2} \bigg[j_{L+ 1}(np_1) j_{L}(np_2) \\
& + j_{L  + 1}(np_2) j_{L}(np_1)\bigg]\sum_{J=0}^{L}  w_J^{1,L} \sum_{S = -J}^{J} Y_{JS}(\hat{p}_1) Y_{JS}^*(\hat{p}_2) \nonumber\\
&= \frac{4}{\pi^2}\sum_{n=0}^{\infty} n\epsilon_n 
\sum_{L = 0}^{\infty} \left[ \phi_{nL}^{2+}(p_1) \phi_{nL}^{2-}(p_2) + \phi_{nL}^{2+}(p_2) \phi_{nL}^{2-}(p_1)\right] \nonumber\\
&\times \sum_{J = 0}^L w_J^{1, L} \sum_{S = -J}^J Y_{JS}(\hat{p}_1) Y_{JS}^*(\hat{p}_2),
\end{align}
with
\begin{align}
&\phi^{2 \pm}_{n L}(p) = p^{1/2} j_{L  + 1/2 \pm 1/2}(np).
\end{align}
$\phi^{2 \pm}_{nL}$ represents the two radial eigenfunctions. Superscript $2$ indicates an inverse-square expansion, $+$ a spherical Bessel function of greater index $(L+1)$ and $-$ one of lesser index $(L)$. 

We can generate all of the other even powers by parametric differentiation with respect to $\cos \theta_{12} \equiv \hat{p}_1\cdot\hat{p}_2$. Expanding the magnitude in the denominator using the binomial theorem we notice that
\begin{align}
&\frac{\partial}{\partial(\cos \theta_{12})} \left[\frac{1}{p_1^2 + p_2^2 + 2p_1 p_2 \cos \theta_{12}}\right] = \frac{2p_1 p_2}{|\vec{p}_1 + \vec{p}_2|^4}
\label{eqn:param_1}
\end{align}
so that we may solve for the desired inverse fourth power as
\begin{align}
&\frac{1}{|\vec{p}_1  + \vec{p}_2|^4} = \frac{1}{2p_1 p_2}\frac{\partial}{\partial(\cos \theta_{12})}\left[ \frac{1}{|\vec{p}_1 + \vec{p}_2|^{2}}\right].
\label{eqn:fourth_ito_square}
\end{align}
We may then insert our representation for $1/|\vec{p}_1 + \vec{p}_2|^2$ in terms of $p_</p_>$ and Gegenbauer polynomials (equation \ref{eqn:gen_exp} with $\lambda = 1$) on the righthand side of equation (\ref{eqn:fourth_ito_square}) and apply the differential operator to just the Gegenbauer polynomials as they are the only factor dependent on $\cos \theta_{12}$. 

We prefer to use parametric differentiation rather than directly changing $\lambda$ from unity to two in equation (\ref{eqn:gen_exp}), as this latter would alter the radial structure of the expansion and so render our particular decoupling integral (\ref{eqn:second_dec_int}) inapt. In contrast, as noted above, parametric differentiation with respect to $\cos \theta_{12}$ acts only on the Gegenbauer polynomial piece of the expansion, preserving the radial structure and hence the utility of our particular decoupling integral. Handled in this way, the radial structure of the expansion for $1/|\vec{p}_1 + \vec{p}_2|^4$ will be the same as for that of $1/|\vec{p}_1 + \vec{p}_2|^2$ save for multiplying by $1/(2p_1 p_2)$.  

Explicitly, we have
\begin{align}
\frac{1}{|\vec{p}_1 + \vec{p}_2|^4} =\frac{1}{2 p_1 p_2} \sum_{L = 0}^{\infty} \frac{p_<^{L}}{p_>^{L + 2\lambda}}\frac{dC_{L}^{(1)}(\cos \theta_{12}) }{d(\cos \theta_{12} )} 
\label{eqn:gen_exp_fourth}
\end{align}
We can use the relation that 
\begin{align}
\frac{d}{dx}\left[ C_{L}^{(\lambda)}(x) \right] = 2 \lambda C_{L - 1}^{(\lambda + 1)}(x)
\label{eqn:gbauer_deriv}
\end{align}
to conveniently evaluate this parametric differentiation while remaining in the basis of Gegenbauer polynomials. This latter is desirable so that we may still apply our ``mixed'' addition theorem (\ref{eqn:C_thm}) splitting the Gegenbauer polynomials into spherical harmonics to factorize the angular dependence. We now simply alter the coefficients $w_J^{L, \lambda}$ to $2w_J^{L - 1, \lambda + 1}$, and the $2$ will cancel with the $1/2$ from equation (\ref{eqn:fourth_ito_square}). We also note that we do not want the sum (\ref{eqn:gen_exp_fourth}) to now involve Gegenbauer polynomials of negative order. Of course, the derivative of the zero-order Gegenbauer polynomial (a constant) vanishes, and so it does not enter the sum. So we leave the indexing of the sums the same but take it that $w_J^{-1,2} = 0$.

We find the factorized eigenfunction expansion for an inverse fourth power as
\begin{align}
\label{eqn:denom_decoup_four}
&\frac{1}{|\vec{p}_1 + \vec{p}_2|^4} = \frac{4}{\pi^2}\sum_{n=0}^{\infty} n\epsilon_n \nonumber\\
&\times \sum_{L = 1}^{\infty} \left[ \phi_{nL}^{4+}(p_1) \phi_{nL}^{4-}(p_2) + \phi_{nL}^{4+}(p_2) \phi_{nL}^{4-}(p_1)\right] \nonumber\\
&\times \sum_{J = 0}^L w_J^{2, L-1} \sum_{S = -J}^J Y_{JS}(\hat{p}_1) Y_{JS}^*(\hat{p}_2),\nonumber\\
&\phi^{4 \pm}_{n L}(p) = p^{-1/2} j_{L  + 1/2 \pm 1/2}(np).
\end{align}
Parallel to our notation in equation (\ref{eqn:denom_decoup_sq}), here superscript $4$ indicates an inverse-fourth-power expansion, $+$ the spherical Bessel function of greater index, and $-$ that of lesser index.

Examining equations (\ref{eqn:denom_decoup_sq}) and (\ref{eqn:denom_decoup_four}) we see that the eigenfunctions may all be conveniently written as
\begin{align}
&\phi_{nL}^{\alpha \pm} (p) = p^{(3 - \alpha)/2} j_{L + 1/2 \pm 1/2}(np)
\end{align}
with $\alpha = 2$ and $4$ for the inverse square and inverse fourth power expansions respectively. This notation enables writing our expansions for both the inverse square and inverse fourth powers as 
\begin{align}
\label{eqn:general_efn}
&\frac{1}{|\vec{p}_1 + \vec{p}_2|^{\alpha} } = \frac{4}{\pi^2} \sum_{n = 0}^{\infty} n \epsilon_n \sum_{L = 0}^{\infty} \bigg[\phi_{nL}^{\alpha +} (p_1) \phi_{nL}^{\alpha - }(p_2)\\
&+   \phi_{nL}^{\alpha +}(p_2) \phi_{nL}^{\alpha - }(p_1)\bigg] 
\sum_{J = 0}^{L} w_J^{\alpha/2, L + 1 -\alpha/2} \sum_{S = -J}^{J} Y_{JS}(\hat{p}_1) Y_{JS}^*(\hat{p}_2),\nonumber 
\end{align}
where we also rewrote the superscripts on the weight $w$ so that they would correctly handle either value of $\alpha$.

Finally, we now write out the full decoupled kernels $K^{\rm d}$ by inserting the expansion (\ref{eqn:general_efn}) into equation (\ref{eqn:Kd_15}). We have
\begin{align}
&K^{\rm d}_{15} =\frac{16}{\pi^4}  \sum_{nn'}W_{nn'} \sum_{L L'}\bigg\{
\phi_{nL}^{\alpha +}(q_1) \phi_{n'L'}^{\alpha' +}(k) \phi_{nL}^{\alpha -}(q_2) \phi_{n'L'}^{\alpha' -}(q_2) \nonumber\\
&+\phi_{nL}^{\alpha +}(q_1) \phi_{n'L'}^{\alpha' -}(k) \phi_{nL}^{\alpha -}(q_2) \phi_{n'L'}^{\alpha' +}(q_2) \nonumber\\
&+\phi_{nL}^{\alpha -}(q_1) \phi_{n'L'}^{\alpha' +}(k) \phi_{nL}^{\alpha +}(q_2) \phi_{n'L'}^{\alpha' -}(q_2) \nonumber\\
&+\phi_{nL}^{\alpha -}(q_1) \phi_{n'L'}^{\alpha' -}(k) \phi_{nL}^{\alpha +}(q_2) \phi_{n'L'}^{\alpha' +}(q_2) \bigg\}\nonumber\\
&\times \sum_{JJ'} \mathcal{W}_{JJ'}^{\alpha \alpha', L L'} \sum_{SS'} Y_{JS}(\hqo) Y_{J'S'}(\hk) Y_{JS}^*(\hqt) Y_{J'S'}^* (\hqt).
\label{eqn:Kd15_mult_out}
\end{align}
Relative to equation (\ref{eqn:Kd_15}) we have chosen to use $\alpha$ in place of $2n_3$ and $\alpha'$ in place of $2n_2'$; $\alpha$ and $\alpha'$ can each take on the values $2$ and $4$ as needed and will point to the correct eigenfunctions (\ref{eqn:denom_decoup_sq}) and (\ref{eqn:denom_decoup_four}). We have also rearranged factors within each term relative to equation (\ref{eqn:Kd_15}) to keep the $\vqt$-dependent factors together. Finally, the sums over $n,\; n', L,$ and $L'$ run from zero to infinity, but the sums over $S$ and $S'$ are bounded by $J$ and $J'$, which are in turn bounded by $L$ and $L'$. The basic structure is the same as that of a spherical harmonic expansion of a Legendre series but then with two additional indices $(n$ and $n')$. In what follows we will not write out the explicit bounds of the sums to save space.

In equation (\ref{eqn:Kd15_mult_out}) we have also defined
\begin{align}
&W_{n n'}  = \epsilon_n \epsilon_{n'} n n',\nonumber\\
&\mathcal{W}_{J J'}^{\alpha \alpha', L L'} = w_J^{\alpha/2, L + 1 - \alpha/2} w_{J'}^{\alpha'/2, L' + 1 - \alpha'/2}.
\label{eqn:W_and_calW_defn}
\end{align}
To keep the notation compact moving forward, we also define the product
\begin{align}
\Pi_{\alpha' \pm, n'L'J'S'}^{\alpha \pm, nLJS} (\vqo, \vk) = \phi_{nL}^{\alpha \pm}(q_1) Y_{JS}(\hqo) \phi_{n' L'}^{\alpha' \pm}(k) Y_{J' S'}(\hk).
\label{eqn:pi_def}
\end{align}
Upper indices correspond to the first argument, lower to the second. We will write $\Pi$ with just one argument when both factors on the righthand side have the same argument, as will be the case for $\Pi$ from the $\vqt$ terms in equation (\ref{eqn:Kd15_mult_out}). With these definitions, we find
\begin{align}
\label{eqn:Kd15_final}
&K_{15}^{\rm d} = \frac{16} {\pi^4} \sum_{n n'} W_{n n'} \sum_{L L' J J'}  \mathcal{W}_{J J'}^{\alpha \alpha', L L'}  \\
&\times\sum_{S S'} \bigg\{\Pi^{\alpha +}_{\alpha'+} (\vqo, \vk) \Pi^{\alpha - *}_{\alpha'-} (\vqt) + \Pi^{\alpha +}_{\alpha'-} (\vqo, \vk) \Pi^{\alpha - *}_{\alpha' + } (\vqt)  \nonumber\\
& + \Pi^{\alpha -}_{\alpha' +} (\vqo, \vk) \Pi^{\alpha + *}_{\alpha'-} (\vqt) + \Pi^{\alpha -}_{\alpha' -} (\vqo, \vk) \Pi^{\alpha + *}_{\alpha' +} (\vqt) \bigg\}.\nonumber
\end{align}
Above we have suppressed the $nLJS,\; n' L' J' S'$ lower and upper indices on $\Pi$ since they are the same for all of the factors in each of the four terms above. $*$ denotes conjugate and should be taken to apply to the entire $\Pi$, not just the part indexed by $\alpha$. As discussed near equation (\ref{eqn:Kd_33}), equation (\ref{eqn:Kd15_mult_out}) also covers the 33 decoupling, as $K^{\rm d}_{33} = K^{\rm d}_{15}$. 

For completeness, we now write out the expression for $K_{24}^{\rm d}$. By comparing equations (\ref{eqn:Kd_15}) and (\ref{eqn:Kd_24}), we see that leaving the $\alpha$ terms unchanged and replacing $\vk$ with $\vqo$ and $\vqt$ with $\vk + \vqt$ in the terms in $\alpha'$ in equation (\ref{eqn:Kd15_final}) will give the desired result. It is
\begin{align}
\label{eqn:Kd24_final}
&K_{24}^{\rm d} = \frac{16} {\pi^4} \sum_{n n'} W_{n n'} \sum_{ L L' J J'}  \mathcal{W}_{J J'}^{\alpha \alpha', L L'} \\
&  \sum_{S S'}  \bigg\{\Pi^{\alpha +}_{\alpha'+} (\vqo) \Pi^{\alpha - *}_{\alpha'-} (\vqt, \vqt + \vk) + \Pi^{\alpha +}_{\alpha'-} (\vqo) \Pi^{\alpha - *}_{\alpha' + } (\vqt, \vqt + \vk)  \nonumber\\
& + \Pi^{\alpha -}_{\alpha' +} (\vqo) \Pi^{\alpha + *}_{\alpha'- } (\vqt, \vqt + \vk) + \Pi^{\alpha -}_{\alpha' -} (\vqo) \Pi^{\alpha + *}_{\alpha' +} (\vqt, \vqt + \vk) \bigg\}.\nonumber
\end{align}
We note that with these replacements, in contrast to equation (\ref{eqn:Kd15_final}), the first factor of $\Pi$ in each term now just has one argument, $\vqo$, while the second factor in each term now has two, $\vqt +\vk$ and $\vqt$. With these factorizations of the $K^{\rm d}$ in hand, we are now ready to return to the full problem of evaluating $\mathcal{I}_{ij, \ell m}$. It is acceptable to have $\vqt + \vk$ in equation (\ref{eqn:Kd24_final}) as this is a convolutional variable for $\mathcal{I}_{24}$.

\section{2-Loop Terms as 3-D Convolutions}
\label{sec:as_convols}
We now have 
\begin{align}
\label{eqn:full_convolution_15}
&\mathcal{I}_{15, \ell m} (k) =\frac{16}{\pi^4} P_{\rm lin} (k)\int \frac{d\Omega_k} {4\pi} \int \frac{d^3 \vqo} {(2\pi)^3} \frac{d^3 \vqt}{(2\pi)^3}\; K_{15}^{\rm d} K_{15}^{\rm c}\\
&= \frac{16}{\pi^4} P_{\rm lin}(k) \sum_{n n'} W_{nn'} \sum_{L L' J J'} \mathcal{W}_{J J'}^{\alpha \alpha', L L'}\sum_{S S'} 
\int \frac{d\Omega_k} {4\pi}\nonumber\\
&\times \bigg\{ T_{15, \ell m, JJ', SS'}^{\alpha \alpha', +  + - -}(n, n'; n_1, n_1', n_2, n_3'; \vk)+ T_{15, \ell m, JJ', SS'}^{\alpha \alpha', +  -  - +}(\cdots) \nonumber\\
& + T_{15, \ell m, JJ', SS'}^{\alpha \alpha', -  + + -}(\cdots) + T_{15, \ell m, JJ', SS'}^{\alpha \alpha',  - - + +}(\cdots) \bigg\}\nonumber
\end{align}
We have defined
\begin{align}
\label{eqn:T_def}
&T_{15, \ell m, JJ', SS'}^{\alpha \alpha', ++--} (n, n'; n_1, n_1', n_2, n_3'; \vk) =\\
&\int \frac{d^3\vqo}{(2\pi)^3}\; \mathcal{R}_{\ell}^{n_1}(q_1) Y_{\ell m}(\hqo) \Pi^{\alpha +}_{\alpha' +} (\vqo, \vk)\mathcal{P}^{n_3'}(\vk + \vqo)  \nonumber\\
&\;\;\times \int \frac{d^3\vqt}{(2\pi)^3}\; \mathcal{R}_{\ell }^{n_2}(q_2)  Y_{\ell m}^*(\hqt) \Pi^{\alpha - *}_{\alpha' -} (\vqt)  \mathcal{P}^{n_3'}(\vqt + (\vk + \vqo)),\nonumber
\end{align}
where $\mathcal{R}$ and $\mathcal{P}$ are defined in equation (\ref{eqn:calRP}). We have omitted the parameters and arguments of $T$ in favor $\cdots$ in all of the $T$ in equation (\ref{eqn:full_convolution_15}) save for the first because they are exactly the same as those in the first. We note that $T$ has all of the power law indices involved in $I_{15}$ in its argument save for $n_2'$ and $n_3$; these instead enter through respectively $\alpha'$ and $\alpha$. $JJ$ and $SS'$ are indices on the $\Pi$ but have been suppressed.

Inserting the more detailed forms for $\Pi$ of equation (\ref{eqn:pi_def}), we can write $T$ in a form that will facilitate analytic evaluation of the angular pieces:
\begin{align}
 &T_{15, \ell m, JJ', SS'}^{\alpha \alpha', ++--} (n n'; n_1, n_1', n_2, n_3'; \vk)= \phi_{n'L'}^{\alpha' +}(k) Y_{J' S'}(\hk)  \nonumber\\
&\times \int \frac{d^3\vqo}{(2\pi)^3}\; \mathcal{R}_{\ell }^{n_1}(q_1) Y_{\ell m}(\hqo)\phi_{nL}^{\alpha+}(q_1) Y_{JS}(\hqo) \mathcal{P}^{n_1'}(\vk + \vqo) \nonumber\\
&\times  I_{15, \ell m, JJ', SS'}^{\alpha \alpha' -- *}(n n'; n_2, n_3'; \vk + \vqo),
\label{eqn:T_with_I}
\end{align}
where we have defined the inner integral
\begin{align}
&I_{15, \ell m, J J', SS'}^{\alpha \alpha' -- *}(n n'; n_2, n_3'; \vk + \vqo) = \nonumber\\
& \int \frac{d^3\vqt}{(2\pi)^3} \; \mathcal{R}_{\ell}^{n_2}(q_2)   Y_{\ell m}^*(\hqt) \phi^{\alpha-}_{nL}(q_2) \nonumber\\
&\times \phi^{\alpha' -}_{n' L'} (q_2) Y^*_{JS} (\hqt) Y^*_{J'S' }(\hqt)\mathcal{P}^{n_3'}(\vqt + (\vk + \vqo)),
\label{eqn:inner_15}
\end{align}
As we see from equations (\ref{eqn:T_def}) through (\ref{eqn:inner_15}), while equation (\ref{eqn:full_convolution_15}) looks involved, each of the four double integrals given by the $T$ is of the same form, and just comes from multiplying the four terms of equation (\ref{eqn:Kd15_final}) with the factors in equation (\ref{eqn:Kc_15}) for $K^{\rm c}_{15}$. Equations (\ref{eqn:full_convolution_15}), and (\ref{eqn:T_with_I}) with the definition (\ref{eqn:inner_15}) are main results of this paper, and display  the $15$ contribution to the 2-loop power spectrum as a sum of nested 3-D convolutions.  

We now write down the analogous results for the $24$ and $33$ contributions. We treat the $33$ contribution first as it is more similar to the $15$ contribution than is the $24$ one.  The only change in going from the $15$ to the $33$ contribution is that what was formerly the $P(k)$ pre-factor now has a different argument (see equation \ref{eqn:w_ij}) and so must come inside the integrals. We have
\begin{align}
\label{eqn:full_convolution_33}
&\mathcal{I}_{33, \ell m} (k) =\frac{16}{\pi^4} \int \frac{d\Omega_k} {4\pi} \int \frac{d^3 \vqo} {(2\pi)^3} \frac{d^3 \vqt}{(2\pi)^3}\; K_{33}^{\rm d} K_{33}^{\rm c}\nonumber\\
&= \frac{16}{\pi^4} \sum_{n n'} W_{nn'} \sum_{L L' J J'} \mathcal{W}_{J J'}^{\alpha \alpha', L L'}\sum_{S S'} 
\int \frac{d\Omega_k} {4\pi}\nonumber\\
&\times \bigg\{ T_{33, \ell m, JJ', SS'}^{\alpha \alpha', +  + - -}(n, n'; n_1, n_1', n_2, n_3'; \vk)+ T_{33, \ell m, JJ', SS'}^{\alpha \alpha', +  -  - +}(\cdots) \nonumber\\
& + T_{33, \ell m, JJ', SS'}^{\alpha \alpha', -  + + -}(\cdots) + T_{33, \ell m, JJ', SS'}^{\alpha \alpha',  - - + +}(\cdots) \bigg\}
\end{align}
with 
\begin{align}
\label{eqn:T33_def}
&T_{33, \ell m, JJ', SS'}^{\alpha \alpha', ++--} (n, n'; n_1, n_1', n_2, n_3'; \vk) =\nonumber\\
&\int \frac{d^3\vqo}{(2\pi)^3}\; \mathcal{R}_{\ell}^{n_1}(q_1) Y_{\ell m}(\hqo) \Pi^{\alpha +}_{\alpha' +} (\vqo, \vk)\mathcal{P}^{n_3'}(\vk + \vqo)  \nonumber\\
&\times  I_{33, \ell m, JJ', SS'}^{\alpha \alpha' -- *}(n n'; n_2, n_3'; \vk + \vqo)
\end{align}
and
\begin{align}
&\mathcal{I}_{33, \ell m, JJ', SS'}^{\alpha \alpha' -- *}(n, n'; n_2, n_3'; \vk + \vqo) =  \nonumber\\
&\int \frac{d^3\vqt}{(2\pi)^3}\; \mathcal{R}_{\ell}^{n_2}(q_2)   Y_{\ell m}^*(\hqt) \phi^{\alpha-}_{nL}(q_2)\phi^{\alpha' -}_{n' L'} (q_2) Y^*_{JS} (\hqt) Y^*_{J'S' }(\hqt) \nonumber\\
&\times \mathcal{P}^{n_3'}(\vqt + (\vk + \vqo)) P_{\rm lin}(|\vqt + (\vk + \vqo)|).
\label{eqn:I33}
\end{align}
We now turn to the 24 contribution. Comparing equations (\ref{eqn:Kd_15}) and (\ref{eqn:Kd_24}), we see that $n_2'$ becomes $n_3'$ in the definition of $\alpha'$, and consequently any $n_3'$ that was written explicitly must now be replaced with $n_2'$. From this we also see that the unprimed $\alpha$ terms are the same, but the primed ones now have $\vk$ becoming $\vqo$ and also $\vqt$ becoming $\vk + \vqt$. And finally from equation (\ref{eqn:w_ij}) the linear power spectrum with argument $\vk + \vqo + \vqt$ in the $33$ integral now has argument $\vk + \vqt$. We thus have
\begin{align}
\label{eqn:full_convolution_24}
&\mathcal{I}_{24, \ell m, JJ', SS'} (k) =\frac{16}{\pi^4} \int \frac{d\Omega_k} {4\pi} \int \frac{d^3 \vqo} {(2\pi)^3} \frac{d^3 \vqt}{(2\pi)^3}\; K_{24}^{\rm d} K_{24}^{\rm c}\nonumber\\
&= \frac{16}{\pi^4} \sum_{n n'} W_{nn'} \sum_{L L' J J'} \mathcal{W}_{J J'}^{\alpha \alpha', L L'}\sum_{S S'} 
\int \frac{d\Omega_k} {4\pi}\nonumber\\
&\times \bigg\{ T_{24, \ell m, JJ', SS'}^{\alpha \alpha', +  + - -}(n, n'; n_1, n_1', n_2, n_2'; \vk) + T_{24, \ell m, JJ', SS'}^{\alpha \alpha', +  -  - +}(\cdots)  \nonumber\\
&+ T_{24, \ell m, JJ', SS'}^{\alpha \alpha', -  + + -}(\cdots) + T_{24, \ell m, JJ', SS'}^{\alpha \alpha',  - - + +}(\cdots) \bigg\}
\end{align}
with 
\begin{align}
\label{eqn:T24_def}
&T_{24, \ell m, JJ', SS'}^{\alpha \alpha', ++--} (n, n'; n_1, n_1', n_2, n_2'; \vk) =\nonumber\\
&\int \frac{d^3\vqo}{(2\pi)^3}\; \mathcal{R}_{\ell}^{n_1}(q_1) Y_{\ell m}(\hqo) \Pi^{\alpha +}_{\alpha' +} (\vqo)\mathcal{P}^{n_1'}(\vk + \vqo)  \nonumber\\
&\times  I_{24, \ell m, JJ', SS'}^{\alpha \alpha' -- *}(n, n'; n_2, n_2'; \vk + \vqo)
\end{align}
and
\begin{align}
&I_{24, \ell m, JJ', SS'}^{\alpha \alpha' -- *}(n, n'; n_2, n_2'; \vk + \vqo) =  \nonumber\\
&\int \frac{d^3\vqt}{(2\pi)^3}\; \mathcal{R}_{\ell}^{n_2}(q_2)   Y_{\ell m}^*(\hqt) \phi^{\alpha-}_{nL}(q_2)Y^*_{JS} (\hqt) Y^*_{J'S' }(\hqt) \nonumber\\
&\times \phi^{\alpha' -}_{n' L'} (|\vqt + \vk|) \mathcal{P}^{n_2'}(\vqt + \vk ) P_{\rm lin}(|\vqt + \vk|).
\end{align}

\section{Evaluating the Convolutions}
\label{sec:evals}
We now evaluate the $T_{ij}$. For each of the cases (15, 33, and 24) we split the work into two steps, the inner convolution first and then the outer. We give detailed derivations for the 15 case and simply quote results for the other two as the steps to follow are analogous.
\subsection{15: Inner convolution}
We first focus on the inner integral over $\vqt$, defined in equation (\ref{eqn:inner_15}). We see that
\begin{align}
& I_{15, \ell m, JJ', SS'}^{\alpha \alpha' -- *}(n,n'; n_2, n_3'; \vk + \vqo) =\nonumber\\
& \left[\mathcal{R}_{\ell }^{n_2}\phi^{\alpha-}_{n L} \phi^{\alpha'-}_{n' L'}Y_{\ell m}^* Y^*_{JS} Y^*_{J'S'} \star \mathcal{P}^{n_3'}  \right](\vk + \vqo)
\label{eqn:I_inner_15}
\end{align}
where star denotes convolution. The convolution is an integration over $\vqt$ but we have suppressed these arguments inside the square brackets and will do so throughout.

To evaluate the convolution, we use the Convolution Theorem to write
\begin{align}
\label{eqn:inner_convolution}
&I^{\alpha \alpha' - - *}_{15, \ell m, JJ', SS'}(n, n'; n_1, n_1', n_2, n_3'; \vec{k}) = \\
&\bigg[ \mathcal{R}_{\ell}^{n_2} \phi_{nL}^{\alpha -}  \phi_{n' L'}^{\alpha' -} Y_{\ell m}^* Y_{JS}^* Y_{J'S'}^* \star \mathcal{P}^{n_3'}\bigg](\vk + \vqo)=\nonumber\\
&{\rm FT}\bigg\{{\rm FT}^{-1}\left\{ \mathcal{R}_{\ell }^{n_2} \phi_{nL}^{\alpha -}  \phi_{n' L'}^{\alpha' -} Y_{\ell m}^* Y_{JS}^* Y_{J'S'}^*\right\} (\vr)\nonumber\\
&\times{\rm FT}^{-1} \left\{ \mathcal{P}^{n_3'}\right\}(\vr)\bigg\}(\vk + \vqo).\nonumber
\end{align}
We have again suppressed $\vqt$, the arguments of the functions being inverse Fourier-Transformed, but for clarity have included $\vr$, the argument the result of the inverse FT will have in configuration space, as well as $\vk + \vqo$, the final argument once we return to Fourier space.

We now evaluate the required inverse and forward FTs. We first treat the more complicated one, the first convolvand in equation (\ref{eqn:inner_convolution}). We have
\begin{align}
\label{eqn:RPhiY_inv_FT}
&{\rm FT}^{-1}\left\{ \mathcal{R}_{\ell }^{n_2} \phi_{nL}^{\alpha -}  \phi_{n' L'}^{\alpha' -}
Y_{\ell m}^* Y_{JS}^* Y_{J'S'}^*  \right\}(\vr)=\\
&\int \frac{d^3 \vqt}{(2\pi)^3}\: e^{-i\vqt\cdot \vr} \mathcal{R}_{\ell }^{n_2}(q_2) q_2^{3/2 - (\alpha + \alpha')/2} j_L(nq_2) j_{L'}(n'q_2)\nonumber\\
&\times Y^*_{\ell m}(\hqt) Y^*_{JS}(\hqt) Y^*_{J'S'}(\hqt) = \nonumber\\
&\sum_{L_2} (-i)^{L_2} \int \frac{q_2^2 dq_2}{2\pi^2} \; \mathcal{J}_{L_2 L L'}^{n n'}(r; q_2) \mathcal{R}_{\ell }^{n_2}(q_2) q_2^{3/2- (\alpha + \alpha')/2}\nonumber\\
&\times \sum_{M_2}Y_{L_2 M_2}(\hr) \int d\Omega_{q_2}\; Y^*_{L_2 M_2}(\hqt) Y^*_{JS}(\hqt) Y^*_{J' S'}(\hqt) Y^*_{\ell m}(\hqt)\nonumber
\end{align}
where we have defined
\begin{align}
\mathcal{J}_{L_2 L  L'}^{n n'}(r; q_2)  = j_{L_2}(q_2 r) j_{L}(nq_2) j_{L' }(n' q_2).
\end{align}
To obtain the second equality we used the plane wave expansion (\ref{eqn:plane_wave_exp}) to expand the complex exponential into spherical Bessel functions and spherical harmonics, leading to the $Y_{L_2 M_2}$  and the $j_{L_2}$ inside the radial integral in equation (\ref{eqn:RPhiY_inv_FT}). 

We see that the angular integral in equation (\ref{eqn:RPhiY_inv_FT}) is just the overlap integral of four spherical harmonics, which can be evaluated in terms of Gaunt integrals, as we show in Appendix \ref{app:math}. We notice that since the result is real, we may take the conjugate of the whole angular integral without altering it, and so we can rewrite it as the integral over four unconjugated spherical harmonics.

Meanwhile, the radial integral over $q_2$ is just a triple spherical Bessel transform of the linear power spectrum weighted by powers of $q_2$ as well as whatever $q_2$ dependence entered the numerator as in equation (\ref{eqn:leg_numer}). We define it as
\begin{align}
\label{eqn:f3_tensor}
&f_{L L' L_2}^{\alpha \alpha' \ell}( n, n'; n_2; r ) =\\
&\int\frac{q_2^2 dq_2}{2\pi^2} \; \mathcal{J}_{L_2 L  L'}^{n n'}(r; q_2)  q_2^{3/2- (\alpha + \alpha')/2}\mathcal{R}_{\ell}^{n_2}(q_2)P_{\rm lin}(q_2),\nonumber
\end{align}
where we take this as the definition of the $f$-tensor on the lefthand side. We note that the minimum power-law weight applied to the linear power spectrum in this tensor is from setting $\alpha$ and $\alpha'$ to their maxima, $4$, and setting $n_2$ to its maximum as well, $2$. In this case we find a weight of $q_2^{-13/2}$, to which we add $2$ to account for the radial Jacobian in spherical coordinates. The power spectrum scales as $q_2$ in the infrared limit, and the spherical Bessel functions scale as $q_2^{L + L' + L_2}$ in this limit.  We thus have a net power of $[2(L+L'+L_2)-7]/2$: the $f$-tensor is infrared-divergent for low orders of the spherical Bessel functions. However, in practice the numerical integrations will be over a finite range, and as long as this is done carefully, these divergences should not affect the end result. Indeed, FFTLog \citep{Hamilton:2000}, one acceleration method for performing these integrals, biases the integrands by power-law weights in any case and so will not see the divergence numerically at all. We will discuss this in more detail in \S\ref{sec:disc}.

Now, the maximum power-law weight in the integrand occurs when $\alpha$ and $\alpha'$ are at their minima, $2$, as is $n_2$, $n_2 = 1$, leading to an overall power law weight of $[2(L+L'+L_2)+1]/2$; this is not divergent in the infrared. We do not concern ourselves with high-$q_2$ (ultraviolet) divergences as we assume that to avoid artifacts from a finite and discretized integration grid, the linear power spectrum will always be multiplied by a smoothing Gaussian, i.e. $\exp[-q_2^2 \sigma^2]$ with $\sigma$$\simeq$$1$ Mpc/$h$.

With the notation outlined above in hand, our full result for the inverse FT of the first, more complicated convolvand in equation (\ref{eqn:inner_convolution}) is
\begin{align}
\label{eqn:hard_FT_final}
&{\rm FT}^{-1}\left\{ \mathcal{R}_{\ell }^{n_2} \phi_{nL}^{\alpha -}  \phi_{n' L'}^{\alpha' -}
Y_{\ell m}^* Y_{JS}^* Y_{J'S'}^* \right\}(\vr)=\\
& \sum_{L_2} (-i)^{L_2} f_{L L' L_2}^{\alpha \alpha' \ell}( n, n'; n_2; r ) \sum_{M_2} \mathcal{H}_{\ell J J' L_2}^{m S S' M_2}Y_{L_2 M_2}(\hr)  .\nonumber
\end{align}
$\mathcal{H}$ is the integral of four spherical harmonics and is defined in equation (\ref{eqn:H_defn}). Noticing that the $f$-tensor is spin-independent, as is the weight $\mathcal{W}$, we see that at the end of the calculation we will be able to sum over the spins $S$ and $S'$ analytically (we recall equation \ref{eqn:full_convolution_15}). This will also prove true of $M_2$ in the end. We will also be able to sum over $J$ and $J'$ analytically because in equation (\ref{eqn:full_convolution_15}) the only other dependence on these is through the weight $\mathcal{W}$ (itself defined in equation \ref{eqn:W_and_calW_defn}). 

We now compute the inverse FT of the second convolvand in equation (\ref{eqn:inner_convolution}).  For $n_3' = 2$ it scales as $1/r$; for $n_3' = 4$ it is formally divergent so we have regularized with an infrared cutoff of the integral at $\epsilon$; we may take the limit $\epsilon \to 0$ at the end of the full calculation. Details are discussed in Appendix \ref{app:simple_kernel_ft}, as well as an alternative regularization scheme. The choice of regularization does not affect the structure of the calculation and so either may be used as dicated by numerical work. We find
\begin{align}
&{\rm FT}^{-1}\left\{\mathcal{P}^{n_3'}\right\}(\vr) \equiv g^{n_3'}(r;\epsilon) \nonumber\\
&= \frac{1}{ 4\pi r},\;\;n_3' = 1;\nonumber\\
&=-\frac{r}{8\pi} + \frac{\cos r\epsilon}{4\pi^2 \epsilon} + \frac{\sin r\epsilon}{4\pi^2 \epsilon^2 r} + \frac{r\;{\rm Si}(r\epsilon)}{4\pi^2},\;\;n_3' = 2,
\label{eqn:gn3}
\end{align}
where ${\rm Si} (x) = \int_{x}^{\infty} dq\; \sin q/q$ is the sine integral. A key aspect of the expressions in equation (\ref{eqn:gn3}) is that neither has any angular dependence, so they will not alter the angular momentum structure of the product in configuration space. As $r\to 0$, $g$ diverges as $1/r$ for $n_3' =1$ but tends to zero for $n_3' = 2$.  

We now write down the product of the two inverse FTs for the inner integral equation (\ref{eqn:inner_convolution}) as
\begin{align}
&I^{\alpha \alpha' - - *}_{15, \ell m, JJ', SS'}(n, n'; n_2, n_3'; \vec{u})=\sum_{L_2} (-i)^{L_2} \mathcal{H}_{L_2 J J' \ell}^{M_2 S S' m}\nonumber\\
&\times {\rm FT}\bigg\{ f_{L L' L_2}^{\alpha \alpha' \ell}(n, n'; n_2; r) g^{n_3'}(r; \epsilon) Y_{L_2 M_2}(\hr) \bigg\}(\vec{u}) \nonumber\\
&= 4\pi \sum_{L_2}  \mathcal{H}_{L_2 J J' \ell}^{M_2 S S' m} Y_{L_2 M_2}(\hat{u}) h_{L L' L_2}^{\alpha \alpha' \ell}(n, n', ; n_2, n_3'; u; \epsilon).
\label{eqn:I15_inner}
\end{align}
We defined
\begin{align}
& h_{L L' L_2}^{\alpha \alpha' \ell}(n, n'; n_2, n_3'; q_1; \epsilon) \nonumber\\
&  = \int r^2 dr\; j_{L_2}(q_1 r) f_{L L' L_2}^{\alpha \alpha' \ell}(n , n'; n_2; r) g^{n_3'}(r; \epsilon)
\label{eqn:little_h}
\end{align}
and also set
\begin{align}
\vec{u} = \vk + \vqo
\end{align}
to simplify notation and facilitate the outer convolution we will perform in the next section (as dicated by equation \ref{eqn:full_convolution_15}). We note that the IR divergence structure of $h$ is given by combing that of $f$ (equation \ref{eqn:f3_tensor}) and $g$ (equation \ref{eqn:gn3}) with the small-argument behavior of the sBF in $h$. We find that $h$ scales as $q_2^{L_2 - 1 + [2(L + L' + L_2) -7]}$ as $q_2 \to 0$, in the worst-case scenario where $f$ and $g$ diverge most strongly.

The other inner integrals, i.e. equation (\ref{eqn:I_inner_15}) but with the superscripted signs changed to $+-$, $-+$, and $++$, are generated by raising either the first index $L$, second index $L'$, or both indices of the $h$-tensor by unity. This is the only change, but for completeness we write out the full expressions below.
\begin{align}
&I^{\alpha \alpha' + - *}_{15, \ell m, LL', JJ', SS'}(n, n'; n_2, n_3'; \vec{u}) =\nonumber\\
&4\pi \sum_{L_2}  \mathcal{H}_{L_2 J J' \ell}^{M_2 S S' m} Y_{L_2 M_2}(\hat{u})  h_{L+1, L', L_2}^{\alpha \alpha' \ell}(n, n', ; n_2, n_3'; u; \epsilon),
\end{align}
\begin{align}
&I^{\alpha \alpha' - +  *}_{15, \ell m, LL', JJ', SS'}(n, n'; n_2, n_3'; \vec{u}) =\nonumber\\
&4\pi \sum_{L_2}  \mathcal{H}_{L_2 J J' \ell}^{M_2 S S' m} Y_{L_2 M_2}(\hat{u})  h_{L, L'+1, L_2}^{\alpha \alpha' \ell}(n, n', ; n_2, n_3'; u; \epsilon),
\end{align}
and
\begin{align}
&I^{\alpha \alpha' + + *}_{15, \ell m, LL', JJ', SS'}(n, n'; n_2, n_3'; \vec{u}) =\nonumber\\
&4\pi \sum_{L_2}  \mathcal{H}_{L_2 J J' \ell}^{M_2 S S' m} Y_{L_2 M_2}(\hat{u}) 
 h_{L+1, L'+1, L_2}^{\alpha \alpha' \ell}(n, n', ; n_2, n_3'; u; \epsilon).
\end{align}

\subsection{15: Outer convolution}
\subsubsection{Core calculation}
We now turn to performing the outer convolution. From equation (\ref{eqn:full_convolution_15}) we have
\begin{align}
\label{eqn:T_outer}
&T^{\alpha \alpha', + + --}_{15, \ell m, JJ', SS'}(n, n'; n_1, n_1', n_2, n_3'; \vk) = \phi_{n' L'}^{\alpha' +}(k)\\
&\times \bigg[\left( \mathcal{R}_{\ell }^{n_1} \phi_{nL}^{\alpha +}  Y_{\ell m} Y_{JS} \right )\star \left(\mathcal{P}^{n_1'} I_{15, \ell m, J J', S S'}^{\alpha \alpha' - - *}(n, n'; n_2, n_3')\right)\bigg] (\vk).\nonumber
\end{align}
We have suppressed the arguments of the convolvands. Using the Convolution Theorem we may write
\begin{align}
\label{eqn:T_eval}
&T^{\alpha \alpha', + + --}_{15, \ell m, JJ', SS'}(n, n'; n_1, n_1', n_2, n_3'; \vk) =\nonumber\\
& \phi_{n' L'}^{\alpha' +}(k){\rm FT}\bigg\{ {\rm FT}^{-1} \left\{ \mathcal{R}_{\ell }^{n_1} Y_{\ell m}Y_{JS} \phi_{nL}^{\alpha +}\right\}(\vec{x})\nonumber\\
&\times{\rm FT}^{-1} \left\{ \mathcal{P}^{n_1'} I_{15,\ell m, J J', S S'}^{\alpha \alpha' --*}(n, n'; n_2, n_3')\right\}(\vec{x})\bigg\}(\vk).
\end{align}
We use $\vec{x}$ as our configuration-space variable above rather than $\vec{r}$ to distinguish from our work on the inner convolution. Evaluating the first inverse FT in equation (\ref{eqn:T_eval}) we obtain
\begin{align}
& {\rm FT}^{-1} \left\{ \mathcal{R}_{\ell }^{n_1} \phi_{nL}^{\alpha +} Y_{\ell m}Y_{JS} \right\}(\vec{x}) =\nonumber\\
&\sum_{L_1 } (-i)^{L_1} f_{L + 1, L_1}^{\alpha \ell}(n; n_1; x) \mathcal{G}_{\ell J L_1}^{m S (-m-S)} Y_{L_1 (-m-S)}^*(\hat{x}).
\label{eqn:first_FT_inv}
\end{align}
$\mathcal{G}$ is a Gaunt integral, defined in equation (\ref{eqn:Gaunt_defn}). We have also defined
\begin{align}
&f_{L_1, L +1}^{\alpha \ell}(n; n_1; x) =\nonumber\\
& \int \frac{q_1^2 dq_1}{2\pi^2} j_{L_1}(q_1 x) \mathcal{R}_{\ell}^{n_1}(q_1) q_1^{3/2 - \alpha/2 } j_{L + 1}(nq_1);
\label{eqn:f2_tensor}
\end{align}
we recall from equation (\ref{eqn:calRP}) that $\mathcal{R}$ contains a linear power spectrum and so the integal above cannot be performed analytically. This $f$-tensor is a limit of the more general three-index one in equation (\ref{eqn:f3_tensor}) with $\alpha'=0$, $L' = 0 = n'$, $n_2 \to n_1$, $L \to L+1$, $L_2 \to L_1$, $r \to x$, $q_2 \to q_1$. Thus the analysis of the divergences presented there can be used here as well.

We need only to treat one additional case, that with a negative sign in the superscript of $\phi$. The result for that case is
\begin{align}
& {\rm FT}^{-1} \left\{ \mathcal{R}_{\ell }^{n_1} Y_{\ell m}\phi_{nL}^{\alpha -}\right\}(\vec{x}) =\nonumber\\
&\sum_{L_1 } (-i)^{L_1} \mathcal{G}_{\ell J L_1}^{m S (-m-S)} Y_{L_1 (-m-S)}^*(\hat{x}) f_{L_1, L }^{\alpha n n_1 \ell}(x),
\end{align}
i.e. only the order of the spherical Bessel function with argument $nq_1$ changes, from $L+1$ to $L$. 

We now obtain the inverse FT of the second term in equation (\ref{eqn:T_eval}). Noticing that the multiplication of $I_{15}$ by $\mathcal{P}^{n_3'}$ does not alter the angular structure of the expansion, we see we will have an inverse FT at the same order sBF as is in $I_{15}$. We find
\begin{align}
\label{eqn:second_FT_inv}
& {\rm FT}^{-1} \bigg\{ \mathcal{P}^{n_1'} I_{15,\ell m, J J', S S'}^{\alpha \alpha' --*}(n, n'; n_2, n_3')
 \bigg\}(\vec{x}) = \\
& 4\pi \sum_{L_2} (-i)^{L_2} H_{L L' L_2}^{\alpha \alpha' \ell}(n, n'; n_2, n_3', n_1'; x) \sum_{M_2}\mathcal{H}_{L_2 J J' \ell}^{M_2 S S' m} Y_{L_2 M_2}(\hat{x});\nonumber
\end{align}
we have suppressed the momentum argument of $I_{15}$, consistent with our earlier convention for inverse FTs, but retain the parameters $n, n',$ etc. for clarity.
We defined
\begin{align}
&H^{\alpha \alpha' \ell}_{L L' L_2}(n, n'; n_2, n_3', n_1'; x) \equiv \nonumber\\
&\int \frac{q_1^2 dq_1}{2\pi^2} j_{L_2}(q_2 x) h_{L L' L_2}^{\alpha \alpha' \ell}(n, n'; n_2, n_3', n_1'; q_1) \mathcal{P}^{n_1'}(q_1).
\label{eqn:big_H}
\end{align}
The IR behavior of $H$ is as that of $h$ with additional weight $q_2^{L_2 + 2(1 - n_1')}$. 

Thus the configuration-space product of the two inverse FTs (equations \ref{eqn:first_FT_inv} and \ref{eqn:second_FT_inv}) required by equation (\ref{eqn:T_eval}) is:
\begin{align}
&\sum_{L_1 L_2} (-i)^{L_1 + L_2} f_{L+1, L_1}^{\alpha \ell} (n; n_1; x) H^{\alpha \alpha' \ell}_{L L' L_2}(n, n'; n_2, n_3', n_1'; x) \nonumber\\
&\times Y_{L_1 (-m-S)}^*(\hat{x}) Y_{L_2 M_2}(\hat{x}) \mathcal{H}_{L_2 J J' \ell}^{M_2 S S' m} \mathcal{G}_{\ell J L_1}^{n S (-m-S)}.
\end{align}
We notice that $L_1$ and $L_2$ are controlled by $\ell, J$, and $J'$, and $J$ and $J'$ are in turn respectively controlled by $L$ and $L'$. These latter two are just the angular momentum indices in our eigenfunction expansions of the two denominators that were expanded into Gegenbauer polynomials. We do not quote the intermediate results for the other combinations of $\phi^{\alpha \pm}$ and $I_{15}^{\alpha \alpha' \pm \pm}$ as these can easily be generated by raising or lowering appropriate angular momentum indices; we will state all results for $T$ at the end of the calculation.

We now take the FT from $\vec{x}$ to $\vec{k}$ as dictated by equation (\ref{eqn:T_eval}), and observe that we will get a third angular momentum and spin, $\mathcal{L}$ and $\mathcal{M}$, from expanding the plane wave entering this FT. However, since we will integrate $T$ over $d\Omega_k$, we will see that $\mathcal{L} \mathcal{M} \to 0 0$ and so $L_2 \to L_2,\; M_2 \to M_1$. But we do not apply this simplication yet. We have for $T$
\begin{align}
\label{eqn:T_15_final}
&T_{15, \ell m, LL', JJ',SS'}^{\alpha \alpha', ++ --}(n, n'; n_1, n_1', n_2, n_3'; \vec{k})  \\
&= 4\pi \phi_{n' L'}^{\alpha' + }(k)\sum_{\mathcal{L}L_1 L_2} i^{\mathcal{L} - L_1 - L_2}\nonumber\\
&\times \int x^2 dx\; f_{L+1, L_1}^{\alpha \ell}(n; n_1; x) H^{\alpha \alpha' \ell}_{L L' L_2}(n, n'; n_2, n_3', n_1'; x) j_{\mathcal{L}}(kx)\nonumber\\
&\sum_{ M_2 \mathcal{M}} (-1)^{(-m-S)} \mathcal{G}_{\mathcal{L} L_1 L_2}^{\mathcal{M} (m+S) M_2} \mathcal{H}_{L_2 J J' \ell}^{M_2 S S' m}\mathcal{G}_{\ell J L_1}^{m S (-m-S)} Y_{\mathcal{L} \mathcal{M}}^*(\hat{k}).\nonumber
\end{align}
The IR behavior of $T$ is as that of $f$ (equation \ref{eqn:f2_tensor}) combined with $H$ (equation \ref{eqn:big_H}) and weighted by an additional $x^{\mathcal{L} + 2}$.

We now need the three other terms, corresponding to the other options for the plus and minus signs in the superscript of $T$. Changes to the innermost two signs affect only $H$, while the outermost signs affect respectively $f_{L+1, L_1}^{\alpha \ell}(n; n_1; x)$ and $\phi_{n' L'}^{\alpha' + }(k)$. Writing out the additional terms entering equation (\ref{eqn:full_convolution_15}) we find
\begin{align}
\label{eqn:T_other_1}
&T_{15, \ell m, LL', JJ',SS'}^{\alpha \alpha', +- -+}(n, n'; n_1, n_1', n_2, n_3'; \vec{k}) = \\
&4\pi \phi_{n' L'}^{\alpha' + }(k)\sum_{\mathcal{L}L_1 L_2} i^{\mathcal{L} - L_1 - L_2}\nonumber\\
&\times \int x^2 dx\; f_{L, L_1}^{\alpha \ell}(n; n_1; x) H^{\alpha \alpha' \ell}_{L, L' + 1, L_2}(n, n'; n_2, n_3', n_1'; x) j_{\mathcal{L}}(kx)\nonumber\\
&\sum_{M_2 \mathcal{M}} (-1)^{(-m-S)} \mathcal{G}_{\mathcal{L} L_1 L_2}^{\mathcal{M} (m+S) M_2} \mathcal{H}_{L_2 J J' \ell}^{M_2 S S' m}\mathcal{G}_{\ell J L_1}^{m S (-m-S)} Y_{\mathcal{L} \mathcal{M}}^*(\hat{k}),\nonumber
\end{align}
\begin{align}
\label{eqn:T_other_2}
&T_{15, \ell m, LL', JJ',SS'}^{\alpha \alpha', -+ +-}(n, n'; n_1, n_1', n_2, n_3'; \vec{k}) = \\
&4\pi \phi_{n' L'}^{\alpha' - }(k)\sum_{\mathcal{L}L_1 L_2} i^{\mathcal{L} - L_1 - L_2}\nonumber\\
&\times \int x^2 dx\; f_{L + 1, L_1}^{\alpha \ell}(n; n_1; x) H^{\alpha \alpha' \ell}_{L+1, L', L_2}(n, n'; n_2, n_3', n_1'; x) j_{\mathcal{L}}(kx)\nonumber\\
&\sum_{M_2 \mathcal{M}} (-1)^{(-m-S)} \mathcal{G}_{\mathcal{L} L_1 L_2}^{\mathcal{M} (m+S) M_2} \mathcal{H}_{L_2 J J' \ell}^{M_2 S S' m} \mathcal{G}_{\ell J L_1}^{m S (-m-S)} Y_{\mathcal{L} \mathcal{M}}^*(\hat{k}),\nonumber
\end{align}
and
\begin{align}
\label{eqn:T_other_3}
&T_{15, \ell m, LL', JJ',SS'}^{\alpha \alpha', -- ++}(n, n'; n_1, n_1', n_2, n_3'; \vec{k}) = \\
&4\pi \phi_{n' L'}^{\alpha' - }(k)\sum_{\mathcal{L}L_1 L_2} i^{\mathcal{L} - L_1 - L_2}\nonumber\\
&\times \int x^2 dx\; f_{L , L_1}^{\alpha \ell}(n; n_1; x) H^{\alpha \alpha' \ell}_{L+1, L' + 1, L_2}(n, n'; n_2, n_3', n_1'; x) j_{\mathcal{L}}(kx)\nonumber\\
&\sum_{M_2 \mathcal{M}} (-1)^{(-m-S)} \mathcal{G}_{\mathcal{L} L_1 L_2}^{\mathcal{M} (m+S) M_2} \mathcal{H}_{L_2 J J' \ell}^{M_2 S S' m}\mathcal{G}_{\ell J L_1}^{m S (-m-S)} Y_{\mathcal{L} \mathcal{M}}^*(\hat{k})\nonumber.
\end{align}
We now integrate against $d\Omega_k/(4\pi)$ (angle-average) as required by equation (\ref{eqn:full_convolution_15}). This cancels the leading $4\pi$ above and sets $\mathcal{L} = 0 = \mathcal{M}$, which in turn requires that $L_1 = L_2$ and $M_1 = M_2$. This brings the relevant Gaunt integral to $(-1)^{M_1}/\sqrt{4\pi}$ (see equation \ref{eqn:Gaunt_defn}) using \cite{NIST_DLMF} \S34.3.1. to evaluate the two 3j-symbols that enter it.\footnote{\url{https://dlmf.nist.gov/34.3}} Denoting the angle-averaged $T$ as $\bar{T}(k)$, we find
\begin{align}
&\bar{T}_{15, \ell m, LL', JJ',SS'}^{\alpha \alpha', -- ++}(n, n'; n_1, n_1', n_2, n_3'; k) = \nonumber\\
&(4\pi)^{-1/2} \phi_{n' L'}^{\alpha' - }(k)\sum_{L_1} (-1)^{L_1}\nonumber\\
&\times \int x^2 dx\; j_0 (kx) f_{L , L_1}^{\alpha \ell}(n; n_1; x) H^{\alpha \alpha' \ell}_{L+1, L' + 1, L_1}(n, n'; n_2, n_3', n_1'; x) \nonumber\\
&\times \mathcal{H}_{L_1 J J' \ell}^{(m+S) S S' m} \mathcal{G}_{\ell J L_1}^{m S (-m-S)}.
\label{eqn:Tbar}
\end{align}
We notice that in equation (\ref{eqn:full_convolution_15}), the only dependence on $S$ and $S'$ is in the $T$, and so we may now perform the sums over these spins. We then may make the definition
\begin{align}
&\omega_{\ell J J' L_1}^{m} \equiv \sum_{SS'}  \mathcal{H}_{L_1 J J' \ell}^{(m+S) S S' m} \mathcal{G}_{\ell J L_1}^{m S (-m-s)}\nonumber\\
&= \sum_{SS'} \sum_{\mathcal{L} \mathcal{M}} (-1)^{\mathcal{M}} \mathcal{G}_{L_1 J \mathcal{L}}^{(m+S) S - \mathcal{M}} \mathcal{G}_{\mathcal{L} J' \ell}^{\mathcal{M} S' m} \mathcal{G}_{\ell J L_1}^{m S (-m - S)}
\label{eqn:small_omega}
\end{align}
where we used the definition of $\mathcal{H}$ equation (\ref{eqn:H_defn}) to obtain the second line; defining $\omega$ is desirable to simplify notation moving forward. We thus have
\begin{align}
\label{eqn:sum_T_spins}
&\sum_{SS'} \bar{T}^{\alpha \alpha', -- ++}_{15, \ell m, LL', JJ', SS'}(n, n'; n_1, n_1', n_2, n_3'; k)=\\ 
&(4\pi)^{-1/2}\phi_{n' L'}^{\alpha' -}(k) \sum_{L_1} (-1)^{L_1} \omega_{\ell J J' L_1}^m\nonumber\\
&\times \int x^2 dx\; j_0(kx) f^{\alpha \ell}_{L, L_1}(n; n_1; x) H^{\alpha \alpha' \ell}_{L + 1, L' + 1, L_1}(n, n'; n_2, n_3', n_1'; x).\nonumber
\end{align}
So we have expressed $T$ in terms of successive transforms against spherical Bessel functions of linear power spectra weighted by power-laws. Returning to equation (\ref{eqn:full_convolution_15}) we see that this means that the full 15 contribution can be written in this way. The expressions for $T$ above are a second major result of this work: they mean that the 15 contribution to the 2-loop power spectrum can be expressed purely as a sum of successive 1-D integral transforms.

\subsubsection{Control of Angular Momenta}
Before going on to address the 33 and 24 contributions, which will have a similar structure, we briefly discuss the structure of our results thus far. We have sums over $n, n', L, L', J', L_1, L_2,$ and $\mathcal{L}$, with $\ell$ as an additional angular momentum. We discuss their provenance in the order they were produced. $\mathcal{\ell}$ came from expanding an arbitrary numerator in our original integral $I_{15}$; in practice these numerators will have very compact support in $\ell$, so the range of $\ell$ is controlled. The sums over $L$ and $L'$ came from our expansion of the coupled denominators into a sum over Gegenbauer polynomials (the radial piece of which was then decoupled using the decoupling integrals). The sum over $J'$ came from the fact that at each $L$, the Gegenbauer polynomial was expanded into a finite number of spherical harmonics. The range of $J'$ at fixed $L$ is thus finite. The sums over $n$ and $n'$ came from our expansion of the decoupling integral into a product of sBFs. 

We now notice that all angular momenta in the problem are controlled by $\ell, L$, and $L'$. Due to the triangle inequality on the 3j-symbol, $L_1$ is controlled to be $|J' - \ell| \leq L_1 \leq |J' + L_1 |$. Returning to equation (\ref{eqn:hard_FT_final}), we see that $\mathcal{H}$ involves $L_2$. The three other angular momenta in $\mathcal{H}$ are controlled at fixed $L$: $\ell$ comes from the numerators, as just discussed, $J=0$ and $J'$ is finite and ranges from $0$ or $1$ up to $L$ or $L'$ depending on $\alpha$ (i.e. whether we expanded an inverse square or inverse fourth power). So $L_2$ is controlled at fixed $L$. Since $L_1$ and $L_2$ are controlled, $\mathcal{L}$ is by the triangle inequality on the 3j-symbols. So we see that deciding at which $L, L', n,$ and $n'$ to truncate our expansion of the denominators we expanded will fully fix the extent of the rest of the sums. Fortunately, \cite{Dominici:2012} find that these expansions converge for a small number of terms, of order less than 10. 

We make three further comments. First, the $f$-tensors are just multiple sBF transforms of the linear power spectrum weighted by power laws; therefore they should be amenable to the Limber approximation as the order of the sBFs grows large.\footnote{The Limber approximation takes it that as their order grows the sBFs become similar to Dirac Delta functions about the point where they first begin to oscillate; \cite{Limber:1953}.} Thus, if going to a large number of terms in the expansion of the denominators were required (it is likely not), it could possibly be done without the computational cost of more radial integrals. This also holds true of our final integral over $j_{\mathcal{L}}(kr)$. 

Second, $\ell$ is fixed by the numerators in the original integral, and should be small, as the 2-loop PT kernels do not have complicated angular structure in their numerators. Meanwhile $J'$ can range up to $L'$, which itself is controlled by our truncation. If we needed to continue to larger $L'$ before truncating for convergence, the Wigner 3-j symbol would become very squeezed, with $J'$$\sim$$L_1 \gg \ell$; these symbols become very small in this limit, scaling as $1/\sqrt{J' + L +1}$, meaning that terms at high $L$ likely contribute rather less to the result. Since $\mathcal{H}$ couples $\ell, J'$ and $L_2$ in the same way as $J'$, $\ell$, and $L_1$ are coupled (since $J=0$, $\mathcal{H}$ reduces to just 3j-symbols), this conclusion holds for $L_2$ as well.

Third, the $f$-tensors may be evaluated using FFTLogs (see \citealt{Hamilton:2000}), as we discuss further in \S\ref{sec:disc}, so the scaling of the integrals becomes $N_{\rm g} \log N_{\rm g}$ rather than $N_{\rm g}^2$, where $N_{\rm g}$ is the number of points used for the $k$ and $r$ integrations each.

\subsection{33: Inner convolution}
We now turn to the 33 contribution to the 2-loop power spectrum. From equation (\ref{eqn:I33}), we read off that we need 
\begin{align}
&I_{33, \ell m, JJ', SS'}^{\alpha \alpha' --*}(n, n'; n_2, n_3'; \vec{k} + \vec{q}_1) = \nonumber\\
& \bigg[ \left(\mathcal{R}_{\ell}^{n_2} \phi_{nL}^{\alpha-} \phi_{n' L'}^{\alpha' -} Y_{\ell m}^* Y_{JS}^* Y_{J' S'}^*\right)\star \left(\mathcal{P}^{n_3'} P_{\rm lin} \right)\bigg](\vec{k} + \vec{q}_1)=\nonumber\\
& {\rm FT} \bigg\{ {\rm FT}^{-1} \bigg\{\mathcal{R}_{\ell}^{n_2} \phi_{nL}^{\alpha-} \phi_{n' L'}^{\alpha' -} Y_{\ell m}^* Y_{JS}^* Y_{J' S'}^*  \bigg\}(\vec{r})\nonumber\\
&\times {\rm FT}^{-1} \bigg\{ \mathcal{P}^{n_3'} P_{\rm lin}\bigg\}(\vec{r})\bigg\}(\vec{k} + \vec{q}_1). 
\end{align}
The first convolvand has the same form as the first convolvand in equation (\ref{eqn:I_inner_15}), so we can simply adopt equation (\ref{eqn:hard_FT_final}) with the definition (\ref{eqn:f3_tensor}) of $f_{L L' L_2}^{\alpha \alpha' \ell}(n, n'; n_2; r)$. The second convolvand is easily evaluated as
\begin{align}
&{\rm FT}^{-1} \bigg\{ \mathcal{P}^{n_3'} P_{\rm lin}\bigg\}(\vec{r}) = \int \frac{d^3\vec{q}_2}{(2\pi)^3}\; e^{-i \vec{q}_2 \cdot \vec{r}} q_2^{-2n_3'} P_{\rm lin}(q_2)=\nonumber\\
& \int_{\epsilon}^{\infty} \frac{q^2 dq_2}{2\pi^2}\; j_0(q_2r) q_2^{-2n_3'} P_{\rm lin}(q_2) \equiv \xi_{\rm lin}(n_3'; r; \epsilon)
\label{eqn:FT_inv_calP}
\end{align}
with the last integral serving to define $\xi_{\rm lin}(n_3'; r; \epsilon)$ and the infrared cut-off $\epsilon$ only genuinely needed for $n_3' = 2$. For  $n_3' = 1$, the integral converges with $\epsilon = 0$, as $P_{\rm lin}(q_2) \propto q_2$ and $j_0 \to 1$ as $q_2 \to 0$, so overall one has $q^2$ as the leading infrared behavior of the integrand. We have termed this result $\xi_{\rm lin}$ becuase it is simply the same transform as used for the linear correlation function, but here weighted by an additional power law specified by $n_3'$.

Using these results we now have
\begin{align}
&I_{33, \ell m, JJ', SS'}^{\alpha \alpha' -- *}(n, n'; n_2, n_3'; \vec{u}) =\sum_{L_2} (-i)^{L_2} \mathcal{H}_{L_2 JJ' \ell}^{M_2 SS' m} \nonumber\\
& \times {\rm FT} \bigg\{ f_{L L' L_2}^{\alpha \alpha' \ell}(n, n'; n_2; r) \xi_{\rm lin} (n_3'; r; \epsilon) Y_{L_2 M_2}(\hat{r}) \bigg\} (\vec{u}) = \nonumber\\
&4\pi \sum_{L_2} \mathcal{H}_{L_2 JJ' \ell}^{M_2 SS' m} Y_{L_2 M_2}(\hat{u}) h_{\xi, LL' L_2}^{\alpha \alpha' \ell}(n, n'; n_2, n_3'; u; \epsilon)
\label{eqn:I_33_in}
\end{align}
with 
\begin{align}
& h_{\xi, LL' L_2}^{\alpha \alpha' \ell}(n, n'; n_2, n_3'; u; \epsilon) \equiv \nonumber\\
&\int r^2 dr\; j_{L_2}(u r) f_{L L' L_2}^{\alpha \alpha' \ell}(n, n'; n_2; r) \xi_{\rm lin}(n_3'; r; \epsilon).
\label{eqn:h_xi}
\end{align}
The above are the analogs of equations (\ref{eqn:I15_inner}) and (\ref{eqn:little_h}), used for the inner convolution in $\mathcal{I}_{15}$.
\subsection{33: Outer convolution}
We now need the outer convolution for $\mathcal{I}_{33}$, the analog of equation (\ref{eqn:T_outer}). From equation (\ref{eqn:T33_def}) we have
\begin{align}
&T_{33, \ell m, JJ', SS'}^{\alpha \alpha', ++ --} (n n'; n_1 n_1', n_2, n_3'; \vec{k} ) = \phi_{n' L'}^{\alpha' +}(k) Y_{J' S'}(\hat{k}) \nonumber\\
&\times \bigg[\left(\mathcal{R}_{\ell}^{n_1} \phi_{nL}^{\alpha +} Y_{\ell m} Y_{JS} \right) \star \left( \mathcal{P}^{n'_1} I_{33, \ell m, JJ', SS'}^{\alpha \alpha' -- *}\right)\bigg] (\vec{k}).
\end{align}
The first and second convolvands are both exactly the same as those in equation (\ref{eqn:T_outer}). For the first convolvand, we may therefore just adopt equation (\ref{eqn:first_FT_inv}); for the second, equation (\ref{eqn:second_FT_inv}). We make a new definition $H_{\xi, L L' L_2}^{\alpha \alpha' \ell}$ as our analog here of equation (\ref{eqn:big_H}): it is just equation (\ref{eqn:big_H}) but with the replacement $h_{LL' L_2}^{\alpha \alpha' \ell} \to h_{\xi, LL' L_2}^{\alpha \alpha' \ell} $ on the righthand side. Our results for the various required $T_{33}$ and $\bar{T}_{33}$ are then exactly as equations (\ref{eqn:T_15_final})-(\ref{eqn:T_other_3}) but with $H \to H_{\xi}$.  For the sake of clarity, we write out explicitly the analog of equation (\ref{eqn:T_15_final}):
\begin{align}
&T_{33, \ell m, LL', JJ',SS'}^{\alpha \alpha', ++ --}(n, n'; n_1, n_1', n_2, n_3'; \vec{k}) = \\
&4\pi \phi_{n' L'}^{\alpha' + }(k)\sum_{\mathcal{L}L_1 L_2} i^{\mathcal{L} - L_1 - L_2}\nonumber\\
&\times \int x^2 dx\; f_{L+1, L_1}^{\alpha \ell}(n; n_1; x) H^{\alpha \alpha' \ell}_{\xi, L L' L_2}(n, n'; n_2, n_3', n_1'; x) j_{\mathcal{L}}(kx)\nonumber\\
&\sum_{M_2 \mathcal{M}} (-1)^{(-m-S)} \mathcal{G}_{\mathcal{L} L_1 L_2}^{\mathcal{M} (m+S) M_2} \mathcal{H}_{L_2 J J' \ell}^{M_2 S S' m}\mathcal{G}_{\ell J L_1}^{m S (-m-S)} Y_{\mathcal{L} \mathcal{M}}^*(\hat{k})\nonumber,
\end{align}
with 
\begin{align}
&H^{\alpha \alpha' \ell}_{\xi, L L' L_2}(n, n'; n_2, n_3', n_1'; x) \equiv \nonumber\\
&\int \frac{q_1^2 dq_1}{2\pi^2} j_{L_2}(q_2 x) h_{\xi, L L' L_2}^{\alpha \alpha' \ell}(n, n'; n_2, n_3'; q_1; \epsilon) \mathcal{P}^{n_1'}(q_1)
\end{align}
with $h_{\xi}$ defined in equation (\ref{eqn:h_xi}).  

Overall, the 33 integrations shared the same angular momentum structure as the 15. The only difference in our work was the addition of one more linear power spectrum inside the integrals; the 15 contribution had two inside in total, but the 33 has three.  This of course means that summing over spins and the simplifications leading to equation (\ref{eqn:sum_T_spins}) proceed exactly in the same manner here, and we find
\begin{align}
&\sum_{SS'} \bar{T}^{\alpha \alpha', -- ++}_{33, \ell m, LL', JJ', SS'}(n, n'; n_1, n_1', n_2, n_3'; k)=\\ 
&(4\pi)^{-1/2}\phi_{n' L'}^{\alpha' -}(k) \sum_{L_1} (-1)^{L_1} \omega_{\ell J J' L_1}^m\nonumber\\
&\times \int x^2 dx\; j_0(kx) f^{\alpha \ell}_{L, L_1}(n; n_1; x) H^{\alpha \alpha' \ell}_{\xi, L + 1, L' + 1, L_1}(n, n'; n_2, n_3', n_1'; x).\nonumber
\end{align}
This shows that $T$ for the 33 contributions to the two-loop power spectrum can be expressed in exactly the same form as those for the 15 contribution, namely as successive transforms  against spherical Bessel functions of linear power spectra weighted by power-laws.

\subsection{24: Inner convolution}
We now obtain the inner convolution for the 24 contribution to the two-loop power spectrum. The angular structure is the same as for the other two contributions, so we may simply adopt equation (\ref{eqn:sum_T_spins}) but with $H \to H_{\xi}$. We find
\begin{align}
&I_{24, \ell m, JJ', SS'}^{\alpha \alpha' -- *}(n, n'; n_2, n_2'; \vec{k}) = \nonumber\\
&\bigg[\left(\mathcal{R}_{\ell}^{n_2} \phi_{nL}^{\alpha -} Y_{\ell m}^* Y_{JS}^* Y_{J' S'}^* \right) \star \left( \phi_{n' L'}^{\alpha' -} \mathcal{P}^{n_2'} P_{\rm lin}\right)\bigg](\vec{r}).
\label{eqn:24_conv}
\end{align}
This convolution is not the same as equation (\ref{eqn:I_inner_15}) but as already noted does have the same angular structure.

For the inverse FT of the first convolvand, we find
\begin{align}
{\rm FT}^{-1}\left\{\mathcal{R}_{\ell}^{n_2} \phi_{nL}^{\alpha -} Y_{\ell m}^* Y_{JS}^* Y_{J' S'}^* \right\} (\vec{r}) &= \sum_{L_2} (-i)^{L_2} f_{L L_2}^{\alpha \ell}(n; n_2; r) \nonumber\\
&\times \sum_{M_2} \mathcal{H}_{\ell JJ' L_2}^{m S S' M_2} Y_{L_2 M_2}(\hat{r}),
\end{align}
with 
\begin{align}
&f_{L L_2}^{\alpha \ell}(n; n_2; r) \equiv \nonumber\\
&\int \frac{q_2^2 dq_2}{2\pi^2}\; j_{L_2}(q_2 r) j_L(nq_2) \mathcal{R}_{\ell}^{n_2} (q_2) q_2^{(3 -\alpha)/2}.
\end{align}
This result is the analog of equation (\ref{eqn:hard_FT_final}) for the 24 contribution; the IR behavior can be read off from the discussion below equation (\ref{eqn:f2_tensor}).

For the second convolvand in equation (\ref{eqn:24_conv}), we have one more factor, $\phi_{n' L'}^{\alpha' -}$, than in the analogous 33 piece (compare equations \ref{eqn:24_conv} and \ref{eqn:FT_inv_calP}). We find
\begin{align}
&{\rm FT}^{-1}\left\{\phi_{n' L'}^{\alpha' -} \mathcal{P}^{n_2'} P_{\rm lin} \right\}(\vec{r}) =\nonumber\\
& \int _{\epsilon}^{\infty}\frac{q_2^2 dq_2}{2\pi^2}\; j_0(q_2 r) q_2^{-2n_2'} P_{\rm lin}(q_2) q_2^{(3-\alpha')/2} j_{L'}(n' q_2)\nonumber\\
&\equiv \xi_{\phi, L'}^{\alpha' -}(n'; n_2'; r; \epsilon).
\end{align}
The IR behavior of $\xi_{\phi}^{\alpha'}$ is as $q_2^{9/2 + L' - 2n_2' - \alpha'/2}$.

By comparing with our work for the 33 contribution, we see that defining $h_{\xi \phi}$ analogously to $h_{\xi}$ of equation (\ref{eqn:h_xi}) but with $\xi \to \xi_{\phi L'}$ and with $f_{L L' L_2} \to f_{L L_2}$ on the righthand side will allow us to adapt equation (\ref{eqn:I_33_in}) to give the result here required. We find
\begin{align}
\label{eqn:I24_in_final}
I_{24, \ell m, JJ', SS'}^{\alpha \alpha' --*} ( n, n'; n_2, n_2'; \vec{k}) &= 4\pi \sum_{L_2} \mathcal{H}_{L_2 J J' \ell}^{M_2 S S' m} Y_{L_2 M_2} (\hat{k}) \\
&\times h_{\xi \phi, L L' L_2}^{\alpha \alpha' \ell}(n, n'; n_2, n_2'; k; \epsilon).\nonumber
\end{align}

\subsection{24: Outer convolution}
We now present the outer convolution and full result for the 24 contribution. The 24 contribution is particularly easy to deal with as the outer and inner convolutions are independent of each other, because the inner convolution is independent of $\vec{q}_1$. From equation (\ref{eqn:T24_def}) the outer convolution is 
\begin{align}
&\bigg[\left(\mathcal{R}_{\ell}^{n_1} \phi_{nL}^{\alpha +} \phi_{n' L'}^{\alpha' +} Y_{\ell m} Y_{JS} Y_{J' S'} \star \mathcal{P}^{n_1'} \right) \bigg](\vec{k})= 4\pi \sum_{L_2} \mathcal{H}_{L_1 J J' \ell}^{(-m-S) S S' m}\nonumber\\
&\times Y_{L_1 (-m-S)}^* (\hat{k}) h_{L + 1, L' + 1, L_1}^{\alpha \alpha' \ell}(n, n'; n_1, n_1'; k; \epsilon).
\label{eqn:24_outer_convol}
\end{align}
To obtain this result, we noticed several parallels with our previous work. First, the first convolvand is just the first term of equation (\ref{eqn:I_inner_15}) with $n_2 \to n_1$, $-- \to ++$, and no conjugates on the spherical harmonics. The lack of conjugates here does not matter as the first convolvand of equation (\ref{eqn:I_inner_15}) and of the above are both real, meaning one can take the conjugate of all spherical harmoincs (the only complex part of either) wihtout altering the result. Second, the second convolvand above is just the second convolvand of equation (\ref{eqn:I_inner_15}) but with $n_3' \to n_1'$. Thus we could adopt equations (\ref{eqn:hard_FT_final}) and (\ref{eqn:gn3}) with these alterations for the two inverse FTs of the two convolvands above, and then use equation (\ref{eqn:I15_inner}) with a conjugate to obtain our ultimate result (\ref{eqn:24_outer_convol}). We simply multiply equation (\ref{eqn:24_outer_convol}) by our result (\ref{eqn:I24_in_final}) for $I_{24,\ell m}$ to obtain $T$. We obtain
\begin{align}
&T^{\alpha \alpha', -- ++}_{24, \ell m, LL', JJ', SS'}(n, n'; n_1, n_1', n_2, n_2'; k) =16\pi^2\nonumber\\
& \times \sum_{L_1 L_2}\sum_{M_2} \mathcal{H}_{L_1 J J' \ell}^{(-m - S) S S' m} \mathcal{H}_{L_2 J J' \ell}^{M_2 S S' m}h_{L + 1, L' + 1, L_1}^{\alpha \alpha' \ell}(n, n'; n_1, n_1'; k; \epsilon) \nonumber\\
&\times h_{\xi \phi, L L' L_2}^{\alpha \alpha' \ell} (n, n'; n_2, n_2'; k; \epsilon) Y^*_{L_1 (-m - S)}(\hat{k}) Y_{L_2 M_2}(\hat{k}).
\end{align}

Integrating over $d\Omega_k/(4\pi)$ sets $L_2 M_2 = L_1 (m+S)$ by orthogonality of the spherical harmonics, eliminating the sums over $L_2$ and $M_2$ (and canceling a factor of $4\pi$). We represent this integrated $T$ as $\bar{T}$. Now summing over spins $S$ and $S'$ we find
\begin{align}
&\sum_{S S'} \bar{T}^{\alpha \alpha', -- ++}_{24, \ell m, LL', JJ', SS'}(n, n'; n_1, n_1', n_2, n_2'; k) =4\pi\nonumber\\
& \times \sum_{L_1 } \Upsilon_{\ell J J' L_1}^m h_{L + 1, L' + 1, L_1}^{\alpha \alpha' \ell}(n, n'; n_1, n_1'; k; \epsilon) \nonumber\\
&\times h_{\xi \phi, L L' L_1}^{\alpha \alpha' \ell} (n, n'; n_2, n_2'; k; \epsilon),
\end{align}
where we defined the angular weight
\begin{align}
\Upsilon_{\ell J J' L_1}^m \equiv \sum_{S S'} \mathcal{H}_{\ell J J' L_1}^{m S S' (-m - S)} \mathcal{H}_{\ell J J' L_1}^{m S S' (m + S)}
\end{align}
by analogy with $\omega_{\ell J J' L_1}^m$ of equation (\ref{eqn:small_omega}). The difference is that $\omega$ was a sum (over spins) of the product of an integral of four spherical harmonics with one of three harmonics, while here we have a sum (over spins) of the product of two integrals over four spherical harmonics. This is a direct result of the structural difference of the 24 contribution from the 15 and 33 contributions; it stems from that fact that for 24, the inner and outer convolutions are actually completely independent.

\section{Discussion}
\label{sec:disc}
Here we discuss a number of possible extensions of this work and other applications of the work. We then make some comments on how a numerical implementation might proceed.

\subsection{Extensions} 
\subsubsection{Incorporating Anisotropy}
We now outline how this work could be extended to redshift space. Redshift-space distortions (RSD) generate a preferred direction (the line of sight), breaking isotropy, and typically multipoles of the power spectrum with respect to the angle to the line of sight, or wedges (bins)  in angle, are used to quantify the anisotropy. Since RSD stem from peculiar velocities, the velocity kernels $G^{(i)}$ also enter the calculation. These have the same form as the density kernels $F^{(i)}$ but with different numerical pre-factors (see e.g. \citealt{Bernardeau:2002}, equations 43-46). Thus there would be two changes to the calculations presented here. 

First, one would replace the density kernels with the appropriate mix of velocity and density kernels; this would affect only the numerator $N$ of equation (\ref{eqn:Ifun}), and that trivially as the constant coefficients would come outside the integrals. Second, one would average over the orientation of the wave-vector $\vec{k}$ at which the power spectrum is measured with some weights, i.e. in equation (\ref{eqn:Ifun}) replace $d\Omega_k \to \mathcal{L}_{\ell}(\hat{k} \cdot \hat{n})$ with $\hat{n}$ the line of sight. This weight can be factored using the spherical harmonic addition theorem (\ref{eqn:sph_add_thm}) to decompose the Legendre polynomial, and the integration over $d\Omega_k$ easily performed. 

We note in particular that due to the azimuthal symmetry about the line of sight that remains even with RSD, the spins do not couple to this change, only the total angular momentum. Thus the simplification of summing $T$ over spins $S$ and $S'$ (e.g. equation \ref{eqn:sum_T_spins}) still holds. We simply first promote $\bar{T}$ of equation (\ref{eqn:Tbar}) to $T_0$, the monopole moment of $T$ with respect to the line of sight, and also seek the higher multipoles $T_{\ell}$ with respect to the line of sight. Alternatively, it is also trivial to obtain wedges by averaging $T$ over bins in $\mu \equiv \hat{k} \cdot \hat{n}$; one just adds a binning function $\Phi(\mu; \bar{\mu}_i)$ in equation (\ref{eqn:Ifun}) as an integration weight rather than the Legendre polynomials, with $\bar{\mu}_i$ denoting the $i^{th}$ angular wedge. 

\subsubsection{Lagrangian Perturbation Theory}
We now briefly turn to consider extending this work to Lagrangian Perturbation Theory (LPT). There is no closed-form expression of order-by-order recursion relations in LPT (\citealt{Bernardeau:2002} \S2.7); however recent work (\citealt{Matsubara:2015}) obtained approximate LPT kernels, and it would be worth considering whether evaluation using them benefits from the technique presented here.

 We note that in the Zeldovich Approximation (ZA) within LPT, the recursion for the density at all orders takes on a particularly simple form where there are only linear momenta in the denominators, and the only appearance of a sum of momenta is in the numerator \citep{Bernardeau:2002}. Thus the technique presented here is likely not needed for the ZA. Nonetheless we observe that the ZA integrals can likely be done as 1-D transforms by performing all of the angular integrations analytically and exploiting the isotropy of the power spectrum, one of the enabling ideas in this work as well.

\subsubsection{Polyspectra}
The techniques of this work could also be used to enable loop calculations of the bispectrum and trispectrum. At a given, fixed order in SPT, the kernels will be the same, just the combination of density fields will be different. But the key ideas here, factorization, decoupling to reduce to convolutions, and then evaluating the angular pieces analytically, should still enable bispectrum and trispectrum to be reduced to 1-D integrals. For instance, see \cite{Simonovic:2018} equation 3.6. for the one-loop bispectrum and \cite{Bertolini:2016hxg} for the trispectrum.  Furthermore, in many modified gravity (MG) models, the kernels are similar to those of SPT (Vernizzi et al. 2018, in prep.); these can thus also likely be dealt with in the same way.

\subsubsection{Further Accelerating Cosmological Parameter Searches}
Derivatives of the power spectrum with respect to the cosmological parameters might be used to further accelerate recomputation of the loop corrections for many different input cosmologies. Clearly, if one Taylor-expanded the linear power spectrum about some fiducial case with respect to variation in the cosmological parameters, one would have fundamental integrals of the derivatives evaluated at the fiducial case. These fundamental integrals would have the same kernels in the integrand as the integrals over the power spectrum computed in this work, as the kernels are cosmology-independent.\footnote{At least for variations close enough to the fiducial cosmology that assuming Einstein-de Sitter remains valid; see \cite{Hivon:1995}.} The variations in the cosmological parameters would sit outside the integrals. One could thus use the techniques outlined in this work to compute the additional required integrals. Similar ideas are discussed in \cite{Fendt:2007}, \cite{Cataneo:2017}, and \cite{Lewandowski:2018}).

\subsection{Other Applications} 
\subsubsection{Redshift-Space 3PCF, Hierarchical 3PCF Model}
The techniques outlined here also have several other LSS applications. First, even in the tree-level redshift-space 3PCF (\citealt{Slepian_RSD_model:2016}, equation 39) one has a term involving $1/k_3^2$, with $k_3 = |\vec{k}_1 + \vec{k}_2|$. There, the integration that occurs when transforming this to the multipole basis in configuration space (of $r_1, r_2$, and Legendre polynomials in the cosine of their enclosed angle $\hat{r}_1 \cdot \hat{r}_2$) was done by introducing a Delta function, rewriting it as the Fourier Transform (FT) of unity, and performing all of the angular integrals.\footnote{See \cite{Szapudi:2004}, \cite{Slepian_RV:2015}, and \cite{Slepian_RSD_model:2016} for development of the multipole basis for the 3PCF.} This procedure resulted in an infinite sum over terms each involving 2-D radial integrals in the end (see equations 16-18). However, using the full decoupling technique presented here, the denominator $1/k_3^2$ could have been decomposed into an eigenfunction expansion, leaving as an end result an infinite sum of 1-D integrals instead. Numerically, the $1/k_3^2$ contribution to the 3PCF multipoles turns out to be small, because of near-exact cancellation of the 2-D integrals in the result of \cite{Slepian_RSD_model:2016} with each other; see discussion below equation 19 of that work. Nonetheless, at the level of precision DESI \citep{DESI:2016} will likely offer on the 3PCF, even small systematic errors in the theoretical modeling should be avoided if possible. More importantly, we suspect that the expansion proposed here for these terms may shed light on why the nearly-exact cancellation making it so small occurs, offering additional insight on the physics of the 3PCF.

Regarding the physics of the 3PCF in the multipole basis, we further note that the Gegenbauer polynomial expansion and decoupling presented here also enables compactly evaluating the ``hierarchical ansatz'' for the 3PCF (\citealt{Groth:1977}) in this basis. This ansatz takes it that the 3PCF $\zeta$ scales as a cyclic sum of products of 2PCFs $\xi$. At large separations $r$ these in turn scale roughly as power laws, $\xi \to 1/r^2$. We thus have, with $\xi_i \equiv \xi(r_i)$,
\begin{align}
&\zeta_{\rm hier.}(r_1, r_2; \hat{r}_1 \cdot \hat{r}_2) \simeq \xi_1 \xi_2 + \xi_2 \xi_3 + \xi_3 \xi_1\nonumber\\
&\simeq\frac{1}{(r_1 r_2)^2} + \left(\frac{1}{r_1^2} + \frac{1}{r_2^2} \right) \frac{1}{|\vec{r}_1 - \vec{r}_2|^2}\nonumber\\
&= \frac{1}{(r_1 r_2)^2} + \left(\frac{1}{r_1^2} + \frac{1}{r_2^2} \right) \frac{4}{\pi^2}\sum_{n=0}^{\infty} n \epsilon_n \sum_{L=0}^{\infty} \bigg[\phi_{nL}^{2+}(r_1) \phi_{nL}^{2-}(r_2) \nonumber\\
&+ {\rm symm.} \bigg]  \sum_{J=0}^L w_{J}^{1, L} \sum_{S = -J}^{J} (-1)^J Y_{JS}(\hat{r}_1) Y_{JS}^*(\hat{r}_2).
\label{eqn:hier_ans}
\end{align}
The phase factor of $(-1)^J$ comes from using parity to convert the spherical harmonic that would have argument $-\hat{r}_2$ into one with argument $\hat{r}_2$ (see equation \ref{eqn:sph_harm_parity}).

We now project $\zeta_{\rm hier.}$ onto the multipole basis by integrating against $(2\ell + 1)/(16\pi^2)$. The factor of $(2\ell + 1)/2$ corrects for the fact that the Legendre polynomials are orthogonal but not orthonormal, and we have an additional $1/(8\pi^2)$ to deal with the redundancy of integrating over $d\Omega_1 d\Omega_2$ when the Legendres only depend on the cosine of the relative angle, $\hat{r}_1\cdot \hat{r}_2$. To perform the integral we expand the Legendre polynomial using the spherical harmonic addition theorem (\ref{eqn:sph_add_thm}) and invoke orthogonality. We thus find the Legendre expansion of the hierarchical-model 3PCF $\zeta_{\rm hier.}$ as 
\begin{align}
&\zeta_{\rm hier.} (r_1, r_2; \hat{r}_1 \cdot \hat{r}_2) = \sum_{\ell =0}^{\infty} \zeta_{{\rm hier.}, \ell} (r_1, r_2)\mathcal{L}_{\ell}(\hat{r}_1 \cdot \hat{r}_2)
\end{align}
with coefficients
\begin{align}
&\zeta_{{\rm hier.}, \ell} (r_1, r_2) = \frac{1} {(r_1 r_2)^2} \delta^{\rm K}_{\ell 0} +  \frac{1 }{\pi^3} \left(\frac{1} {r_1^2} + \frac{1} {r_2^2} \right)(-1)^{\ell}(2\ell + 1) \nonumber\\
&\times \sum_{L = 0}^{\infty} w_{\ell}^{1, L} \sum_{n = 0}^{\infty} n \epsilon_n\bigg[\phi_{nL}^{2+}(r_1) \phi_{nL}^{2-}(r_2) + {\rm symm.}\bigg].
\label{eqn:hier_coeffs}
\end{align}
$\delta^{\rm K}$ is the Kronecker delta, unity when its subscripts are equal and zero otherwise.

Equation (\ref{eqn:hier_coeffs}) is thus a compact expression for a toy-model of the 3PCF in the multipole basis based on the hierarchical ansatz plus the large-scale behavior of the galaxy correlation function. This toy model could be of use in understanding the scale dependence of $\ell > 0$ multipoles of the 3PCF in the space of $r_1$ and $r_2$, such as are shown in e.g. Figure 9 of \cite{Slepian_RV:2015} or Figure 7 of \cite{Slepian_RSD_model:2016}.  

\subsubsection{An $\mathcal{O}(N)$ $N$-body Integrator}
Another application of the decoupling approach presented here is to $N$-body calculations. Naive evaluation of the gravitational forces a given set of $N$ particles (i.e. a discretization of the matter field) exert on each other scales as $N^2$. Most currently standard algorithms use approximate methods to avoid this scaling. For instance, one might compute forces from nearby particles directly using a pair count but farther away particles using a multipole expansion of their spatial coordinates (e.g. as \textsc{Abacus} does; Garrison et al. 2018, in prep.). However one must recompute the multipole expansion about every particle on which one wants the force (although Abacus uses the spherical harmonic shift theorem to mitigate this cost). Alternatively, one might compute the far-field forces using fast Fourier transforms (Ewald summation). This latter scales as $\Ng \log \Ng$ with $\Ng$ the number of grid points used for the FFT. 

The eigenfunction expansion developed in this work offers an exact expansion of the force law between two particles in absolute coordinates. This latter point is key. By using the decoupling integral and then the integral-to-sum identity to derive this eigenfunction expansion, we removed the restriction that one distance from the global origin be less than another. Thus one can work always in an absolute coordinate system and never need to compare the distances of two particles from the global origin. This avoids a pair count. 

In more detail, here is how an $N$-body algorithm based on this decoupling would proceed. Setting Newton's constant $G=1$ and assuming that all particles have the same mass, we have the acceleration of the $i^{th}$ particle as 
\begin{align}
\vec{a}_{i} = \sum_{j \neq i} \frac{\vec{r}_j - \vec{r}_i}{|\vec{r}_j - \vec{r}_i|^3}.
\end{align}
With an eigenfunction expansion of the denominator, in this case in Legendre polynomials using parametric differentiation as discussed in \S\ref{sec:decoupling}, around equations (\ref{eqn:Dominici_decomp}), (\ref{eqn:param_1}) and (\ref{eqn:fourth_ito_square}) (for the Gegenbauer case, but the same idea can be used to obtain an inverse-cube from the Legendre series), we have
\begin{align}
&\vec{a}_{i} = \sum_{j \neq i} (\vec{r}_j - \vec{r}_i) \sum_{nL}\sum_{n'L'} \phi_{nL}^{[3]}(r_i)\phi_{n'L'}^{[3]}(r_j) \nonumber\\
&\times \sum_{JJ'} w^{L}_{J} w_{J'}^{L'} \sum_{SS'} Y_{JS}(\hat{r}_i) Y_{J'S'}(\hat{r}_j).
\end{align}
$J$ and $J'$ run from zero up to $L$ and $L'$ respectively and come from re-expressing the derivative of a Legendre polynomial as a series in lower-order ones with weights $w_{J}^{L}$ and $w_{J'}^{L'}$. These Legendres are then expanded into spherical harmonics. We may now rewrite the sum that excluded $i$ as one including $i$ and then subtract the value of the summand at $i$. 

However we assume that the truncation of our series in $n, n', L$, and $L'$ will offer a natural softening so that the force kernel does not diverge as $\vec{r}_j \to \vec{r}_i$. In particular, Figure 1 of \cite{Dominici:2012} shows that the sum approximating the integral, when truncated, does not develop a singularity at equal values of the free frequencies ($a$ and $b$ in that work, $r_i$ and $r_j$ here), at least for the orders of sBF they tested. This point is akin to the idea that if one expands a Dirac delta function in Fourier modes but limits the bandwidth, one will recover an oscillatory, smoothed function with infinite support. Given Parseval's theorem, that the integral of a function is conserved from configuration space to Fourier space, the broadening implies that the singularity of the Delta function at the origin must now be finite. Consequently we suspect that truncating the sums in the eigenfunction expansion, as is required to compute them numerically in any case, will offer a ``natural'' softening.\footnote{Given that softening has recently been found to cause problems in $N$-body simulations \citep{vandenB:2018}, it may be of interest in and of itself to investigate whether the ``natural'' softening proposed here offers improvements over standard softenings.}  If this holds, then the value of the summand at $j=i$ will tend to zero, so that we can drop that term. We then find
\begin{align}
&\vec{a}_{i} =  \sum_{nLJS} \phi_{nL}^{[3]}(r_i) w^{L}_{J} Y_{JS}(\hat{r}_i)\sum_{jn'L'J'S'} \vec{r}_j  \phi_{n'L'}^{[3]}(r_j)w^{L'}_{J'}  Y_{J'S'}(\hat{r}_j) \nonumber\\
& - \vec{r}_i  \sum_{nLJS} \phi_{nL}^{[3]}(r_i) w^{L}_{J} Y_{JS}(\hat{r}_i)\sum_{jn'L'J'S'}  \phi_{n'L'}^{[3]}(r_j) w^{L'}_{J'} Y_{J'S'}(\hat{r}_j).
\end{align}
We now notice that, at fixed $i$, the two sums depending on $i$ are just numerical values, and can be pre-computed for every particle in the simulation box (i.e. stored for each $i$). The sums over $j$ are both over all particles in the box, and can therefore just be computed once. Furthermore, once the eigenfunctions $\phi$ and $Y_{JS}$ are found at all particle positions, these can be used to quickly perform both of these computations. The acceleration of the $i^{th}$ particle in the box at a given time-step is then just found using the simple linear combination of products these results as given above.

\subsection{Comments on Numerical Implementation}
We defer a full numerical implementation to future work, although it should be straightforward save for requiring some care to track all indices without error. However, we outline a few ideas on how an efficient numerical implementation might proceed.  

We first sketch how even the most naive scheme could be rather efficient. Each successive transform required for our final results is just an integral of the power spectrum weighted by power laws against one or more spherical Bessel functions. This is exactly the type of integral done in CMBFAST (\citealt{Seljak:1996}) to compute predictions for the temperature and polarization anisotropies of the Cosmic Microwave Background (CMB). There, the integration is split into a source function and geometric part, the source function being the perturbation theory potentials (for instance $h$ and $\eta$ in synchronous gauge, see e.g. \citealt{Ma:1995}). The geometric part is the spherical Bessel functions, which can be pre-computed into a lookup table, meaning they need only be obtained once. Since the source functions are smooth but the spherical Bessel functions are oscillatory, these latter drive the cost of sampling and evaluation. However since the geometric part is known and does not change with changing source functions, one can optimize and precompute the sampling points for the integrations. This point is discussed in the CMBFAST implementation (\citealt{Seljak:1996}), and there only about 50 points are required for the integrations.

In our case, the power spectrum plays the role of source function, and even despite BAO it will be rather smooth relative to the highly oscillatory sBFs. Thus in our case as well the sBFs will drive the integration points. A further advantage of our decomposition is that many of the spherical Bessel functions appearing in our expansions have integer frequencies; this makes locating the nodes especially easy, likely enabling further optimization of the sampling for the numerical integration. We note that the most naive integration scheme would scale as $\Ng^2$ with $\Ng$ grid points for the momenta $q_i$ and $k$ and the same for the configuration space-variables $r$ and $x$. At each momentum $q_i$ or $k$, one would need to evaluate an integral over all $r$ or $x$. 

However, these integrals can be done scaling as $\Ng \log \Ng$ using a generalization of the FFTLog algorithm of \cite{Hamilton:2000}, as mentioned briefly at the end of \S5.2. The change of variable outlined there renders integrals against a single spherical Bessel convolutional, and this change of variable is actually agnostic as to how many sBFs one has, so can also be used to render integrals against multiple sBFs so. These ideas are further discussed in \cite{Li:2018} and Li and Slepian 2018 (in prep.).\footnote{\url{https://github.com/eelregit/mcfit}} There are two advantages of this approach, in addition to the speed increase. First, it naturally handles logarithmically gridded input kernels, such as one would typically have for a power spectrum formed from transfer functions from e.g. CMBFAST \citep{Seljak:1996}, CAMB \citep{Lewis:2000}, or CLASS \citep{Lesgourgues:2011}. Second, the FFTLog transformation ``biases'' the integrands by power laws to take out any smooth, secular fall-off or rise and thereby increase the numerical precision of the integration. \cite{Hamilton:2000} shows that this biasing can then be undone analytically in the final space. So FFTLog will not actually ``see'' divergences in the integrals caused by power-law weights, which are exactly the divergences that arise in the approach of this work.

\begin{figure}
\centering
    \includegraphics[width=.48\textwidth]{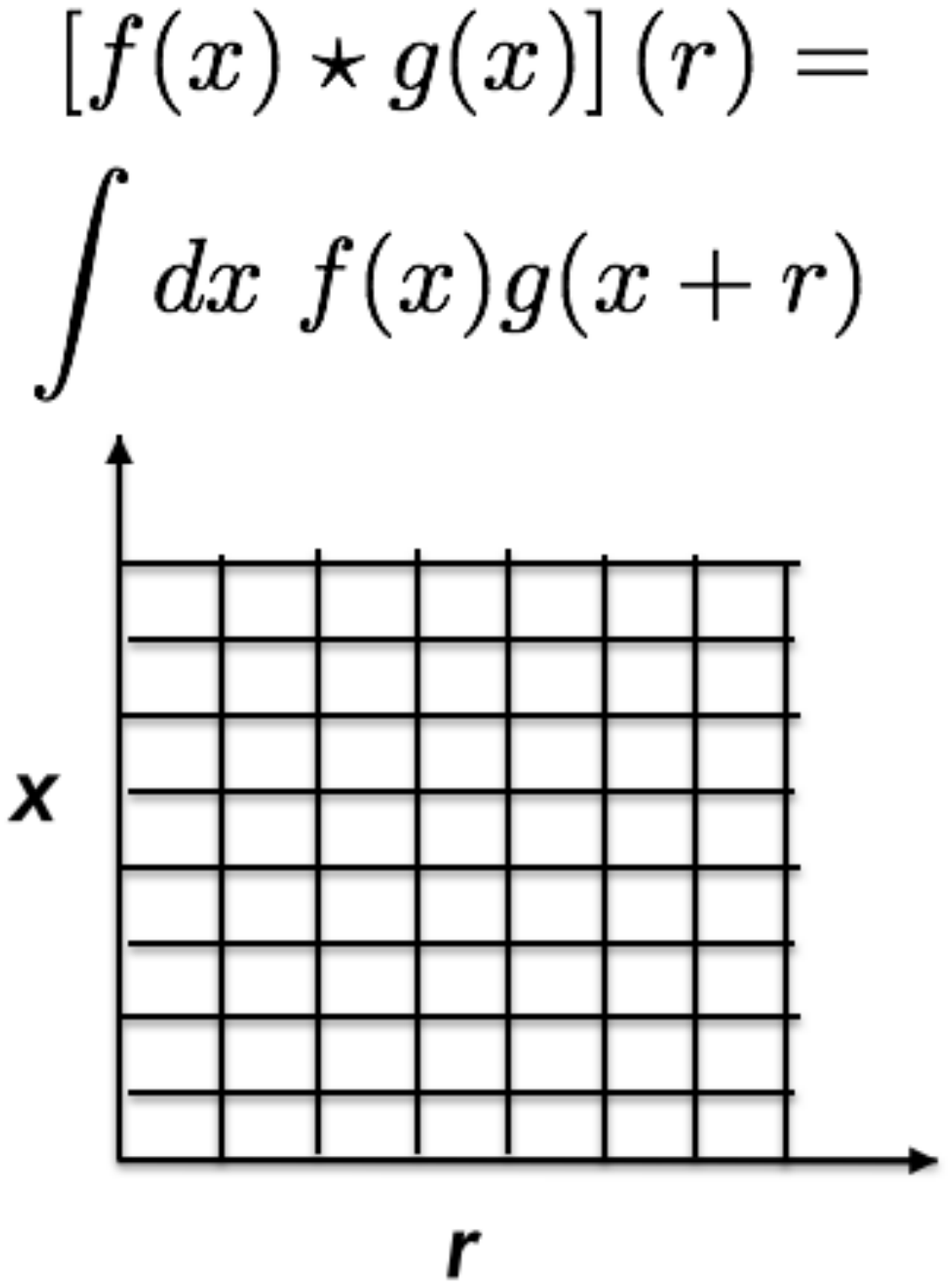}
	\caption{Naively a convolution requires significant compuational work. For instance a 1-D convolution requires a 2-D integration grid, as the diagram above illustrates. At each $r$ one must integrate over all $x$. The scalng is thus $\Ng^2$ with $\Ng$ the number of grid points in $x$ and $r$. However, the advantage of using FFTs is that the Convolution Theorem states one can mutliply the FFTs and then inverse FFT; the cost of the multiply is $\Ng$, while the cost of the FFTs is $\Ng \log \Ng$ due to the existence of very efficient FFT algorithms.}
\label{fig:convol_cost}
\end{figure}

\section{Connections to Other Work}
Here we discuss connections to other work, in particular highlighting the power-law-basis approach to these integrals and also highlighting similar techniques used in QFT.
\subsection{Complex Power-Law Basis}
We first discuss a possible connection with the power-law work of \cite{Simonovic:2018}. There, the power spectra are expanded as a sum of complex power laws (the imaginary part of the power laws enables capturing oscillations, such as BAO) and the integrals of the full PT kernels against complex power laws are done analytically. Thus when one alters the input power spectrum one only needs to update the expansion coefficients into the complex-power-law basis, and can then sum up the integrals (done already in closed form) with these new weights. The same approach could be applied here: writing $P_{\rm lin}$ as a sum of complex power laws would allow analytic evaluation of the 1-D integrals in our final results. 

The advantage of this ``hybrid'' approach over that presented in \cite{Simonovic:2018} is that one would likely have easier results to evaluate than the hypergeometric functions required by their analytic integrals. The disadvantage is that some of these might diverge and some care might be required in treatment so as to obtain a convergent sum in all cases, and also that one would have several infinite (but countably so) sums. Details may merit attention in future work. However, taking that aside, we note that the formulae derived here must, by comparison with the results of \cite{Simonovic:2018}, offer a new expansion of the hypergeometric functions into infinite sums of likely relatively simple terms. In particular, as noted above, mating our approach with a complex-power law expansion of the power spectrum, one would end up with integrals of complex power laws against pairs or triplets of spherical Bessel functions in place of the hypergeometrics appearing in \cite{Simonovic:2018}. One can do these former in simple form, including resolution of the singular terms, as shown in Slepian (2018, in prep.) and Cahn and Slepian (2018, in prep.).

\subsection{Multi-Loop Calculations in Quantum Field Theory}
\label{subsec:QFT}
The present work also is related to techniques for performing multi-loop calculations in QFT. In particular, the first part of our approach here---expansion of certain denominators into a Gegenbauer series---is exactly what is done in the Gegenbauer Polynomial $x$-space Technique (GPXT) developed in \cite{Chetyrkin:1980} and originally used to compute the first 3-loop results for $e^{+} e^{-} \to$ hadrons. GPXT was further developed in \cite{Chetyrkin:1979}, \cite{Celmaster:1980}, \cite{Terrano:1980}, with some early calculations presented in \cite{Chetyrkin:1984}, \cite{Lampe:1983}, \cite{Smirnov:1991}, \cite{Chetyrkin:1993}, and \cite{Kotikov:1996}. However in QFT the Feynman integrals have no source functions in the integrand that are known solely numerically; all the functions in the integrand are written down in closed form. Thus the momentum-magnitude integrals may reduce to a set of nested integrals on a simplex (e.g. equations 2.8 and 2.9 of \citealt{Chetyrkin:1980}), which can then be done just once analytically. In our case, where one may want to use many different linear power spectra, this would be inefficient. Avoiding the inefficiency of working over simplexes is where the utility of our decoupling integral and integral-to-sum identity lies. 

We note that our approach does likely simplify some QFT integrals even though the integrand is completely known in closed form. For instance, in \cite{Bekavec:2006} GPXT is applied to a 5-loop calculation, leading to a result that has 120 different terms corresponding to the $5!$ different simplexes one needs to consider for the different orderings of the position-space lengths (Fourier dual of the momentum magnitudes) being integrated over. Although due to symmetries only 30 terms are genuinely different, this is still a substantial number. The eigenfunction expansion presented here, by breaking the simplicial coupling, enables avoiding individual treatment of each simplex, potentially reducing the computational complexity.

There are several other works in QFT that seem conceptually related to the present one. \cite{Bollini:1996} exploits the spherical symmetry of the Feynman propagators in Fourier space to rewrite them as $j_0$ transforms of configuration space functions, and then performs the convolution (e.g. of two) as a product in configuration space. This product gives two sBFs. Exploiting the spherical symmetry of the result and transforming back to Fourier space yields a third. The transform to Fourier space also gives an integration over the configuration space variable: this can be performed analytically as the overlap integral of the three sBFs. One then has integrals left over for the two initial $j_0$ transforms now weighted by a coupling kernel from the triple-sBF overlap integral. This coupling kernel has a geometric interpretation: it is proportional to the area of a triangle whose sides are the external momentum and these two internal momenta. This work is related insofar as it exploits spherical symmetry and therefore works in terms of sBFs; however the result, of a coupled integral over momenta, is exactly what we here wished to avoid as this would couple the integrals over power spectra that must be done numerically, leading to a 2-D or more grid for numerical integration.

An interesting connection to the technique above is work presented in \cite{Davydychev:1997} showing that 1-loop Feynman integrals of $N$ propagators can be related to the volumes of $N$-dimensional simplexes. \cite{Ortner:1995} also relates this type of integral to simplexes. The connection to the technique above is the appearance there of the area of the triangle between the external and two internal momenta: in essence the 2-D volume of a simplex. It is not surprising that simplexes show up in these works: as we have shown here, expansion of the integrands into Gegenbauer series enforces ``simplicial'' constraints between the momentum magnitudes. 

Our work here is also realted to that of \cite{Easther:2000}, which expands the Feynman propagator in a sum of generalized sinc functions. This expansion converges quickly, and also renders all of the integrals one needs to perform Gaussian, allowing most of the calculation to be done analytically just once. This is conceptually similar to \cite{Simonovic:2018}, which also exploits the opportunity to perform integrals over a set of basis functions analytically.

Finally, regarding QFT, we note that our technique, as a method of handling simplexes, can also be used as an alternative representation of the time-ordering operator $\hat{T}$ that appears in Dyson series. In particular, our decoupling integral (reference equation) for the even powers, when not symmetrized, is identically zero when the frequency of the smaller-order sBF exceeds that of the larger-order sBF. This can be used to rewrite the time-ordered product that enters the integrals giving the Dyson series. Thus one obtains a purely algebraic representation of $\hat{T}$. Whether this has utility in performing QFT calculations with Dyson series, or is simply of conceptual interest, may be worth exploring in future work.

\section{Conclusion}
\label{sec:concs}
In this paper, we have shown that the integrals required for the 2-loop power spectrum in cosmological Eulerian Standard Perturbation Theory can be written as nested double 3-D convolution. The kernels one integrates against linear power spectra to generate the 2-loop corrections involve six different denominators each containing dummy momenta and possibly the external momentum. Four of these denominators can be automatically handled as convolutions, with the particular choice of four dictated by the arguments of the linear power spectra. The remaining two need to be decoupled so that they can be grouped into the convolvands entering each of the nested convolutions. 

We accomplish this decoupling in four steps. First, we write each of the denominators as a Gegenbauer series; Gegenbauer polynomials in the dot product of the relevant unit vectors appear because the denominators all involve even powers (two and four). This factorizes the radial and angular dependences from each other. Second, we decouple the dot product in the Gegenabauers' arguments by expanding each Gegenbauer into a sum of spherical harmonics depending on one unit vector each. Third, we decouple the radial piece, which is formally expanded in powers of the ratio of the greater to the lesser magnitude vector, by rewriting it as an overlap integral of two spherical Bessel functions. This integral automatically selects the correct (smaller) magnitude to be in the numerator of the ratio. However, to avoid paying the price of performing an integral at the end of our convolutions, which would require the convolutions themselves to be of higher dimension, we use an integral-to-sum identity to rewrite the overlap integral as an infinite series, though one that quickly converges. These steps give a complete decoupling and factorization of the 2-loop integrals into nested, double 3-D convolution. Since all of the angular dependence is known in closed form (as the power spectra are isotropic), we perform all of the angular integrations in each 3-D convolution to obtain 1-D radial integrals. We then have expressions for the 2-loop corrections that simply involve a sequence of 1-D integrations. 

This approach is related to the long-used QFT multi-loop calculation aid Gegenbauer Polynomial $x$-Space Technique (GPXT), although this latter does not decouple the radial piece appearing in the Gegenbauer series. This work solves the same problem, but in a different way, as recent work by \cite{Simonovic:2018}. That work expands each linear power spectrum as a series in complex power laws, and performs the integrals of the SPT kernels against these basis functions analytically, resulting in hypergeometric functions. The computational cost of that method is therefore front-loaded; initial calculation of the hypergeometric functions to high precision and with appropriate analytic continuations is challenging, but once this is done, only one FFT is required to obtain the complex power law coefficients of the linear power spectrum, and this vector can then be matrix-multiplied by the basis integrals. 

In contrast, in the approach of the present work, no pre-computation is done, and one simply has a series of forward and inverse FFTs of the linear power spectra. The exact number of FFTs required will depend on the order to which both the Gegenbauer series and the series for the integral-to-sum identity used to handle the sBF overlap integral. The number of integrals could likely be somewhat reduced using recursion relations---taking a linear combination of two sbF integrals with a given power law weight could provide another sBF integral with a different power-law weight. More detailed comparison of the computational cost of the present method with that of \cite{Simonovic:2018}, in particular the exact balance of less initial cost but more run-time calculation, is probably best done once the present method has been numerically implemented. 

We have also shown that a few desirable extensions, such as to redshift-space and to polyspectra (bispectrum, trispectrum), should be straightforward. Further, we have presented several additional applications of the technique outlined here, such as its possible utility for simplifying multi-loop integrals in QFT and its potential to serve as a basis for a novel $N$-body integration scheme. 

More generally, we note that the importance of factorization into radial and angular pieces as well as the appearance of sBFs and spherical harmonics is not surprising. Given the isotropy of the power spectrum, this factorization makes sense, and many other works exploit this framework as well. For instance, \cite{Dai:2012} and \cite{Dai:2013} work in a total angular momentum basis where they expand the observables using spherical harmonics and their generalizations motivated by the spherical symmetry in cosmology. Earlier work by \cite{Zaroubi:1993} displayed cosmological SPT to second-order in the basis of spherical harmonics. 

Looking forward, we expect that techniques such as that outlined here will be useful in the context of large-volume, high-precision surveys such as DESI or LSST. DESI especially has BAO as one major focus, and BAO scales are large enough that perturbative methods are valid. However, DESI will achieve percent precision on of order ten redshift bins, for sub-percent precision overall, and so accurately including higher-order PT corrections will be worthwhile. Furthermore, given the high precision of DESI, it is likely that MCMC exploration of variations in the cosmological parameters will also be rewarded. This will require recomputation of the loop corrections a large number of times for different input linear power spectra. Thus the speed-up in loop calculations offered by the method presented here (or that of \citealt{Simonovic:2018}) will be particularly enabling.

\section*{Acknowledgements}
ZS thanks Emanuele Castorina, Jolyon K. Bloomfield, Joanne D. Cohn, Alex Krolewski, Daniel J. Eisenstein, Douglas P. Finkbeiner, Yin Li, Stephen Portillo, Marcel Schmittfull, Marko Simonovi\'c, Masahiro Takada, and Martin White for useful conversations, as well as Cornelius Rampf and Marko Simonovi\'c for comments on a draft. ZS especially expresses gratitude to Bob Cahn for numerous discussions and encouragement over the long period of this work as well as comments on many calculations. ZS also thanks Pat McDonald and Jolyon Bloomfield for each sharing unpublished notes. ZS particularly thanks Donghui Jeong, Sarah Shandera, Matias Zaldarriaga, and Matt McQuinn for hospitality at respectively the Pennsylvania State University, the Institute for Advanced Study, and the University of Washington during portions of this work, as well as Joshua Silver, Roberta Cohen, Hillary Child, and Salman Habib for the same in Chicago and at Argonne National Laboratory. Support for portions of this work was provided by the National Aeronautics and Space Administration through Einstein Postdoctoral Fellowship Award Number PF7-180167 issued by the Chandra X-ray Observatory Center, which is operated by the Smithsonian Astrophysical Observatory for and on behalf of the National Aeronautics Space Administration under contract NAS8-03060. ZS also received support during portions of this work from the Berkeley Center for Cosmological Physics as well as affiliate resources from Lawrence Berkeley National Laboratory.




\bibliographystyle{mnras}
\bibliography{pt_decoup_refs} 




\appendix
\section{Relevant Mathematical Theorems and Identities}
\label{app:math}
The spherical harmonic addition theorem decouples Legendre polynomials $\mathcal{L}_{\ell}$ whose argument is a dot product of two unit vectors into a sum of products of spherical harmonics $Y_{\ell m}$ each in just one unit vector, as (e.g. \citealt{AWH13})
\begin{align}
\mathcal{L}_{\ell}(\hat{x} \cdot \hat{y}) = \frac{4\pi}{2\ell + 1} \sum_{m = -\ell}^{\ell}Y_{\ell m}(\hat{x})Y^*_{\ell m}(\hat{y}).
\label{eqn:sph_add_thm}
\end{align}
The spherical harmonics we use are complex:
\begin{align}
Y_{\ell m }(\hat{r} )= \sqrt{\frac{4\pi}{2\ell + 1}}P_{\ell}^m(\cos\theta)e^{im\phi}
\end{align}
where $\theta$ and $\phi$ are respectively the angle of $\hat{r}$ with the $z$-axis and the azimuthal angle of $\hat{r}$ with respect to the $x$-axis (i.e. we take the $x$-axis as our zero point for $\phi$). $\phi$ runs from $0$ to $2\pi$ and $\theta$ from $0$ to $\pi$. $P_{\ell}^m$ is an associated Legendre polynomial.

The conjugate identity for spherical harmonics changes the sign of $m$ and gives a pre-factor of $(-1)^m$, i.e.
\begin{align}
Y_{\ell m}^*(\hat{r}) = (-1)^m Y_{\ell -m}(\hat{r}).
\label{eqn:conj_id}
\end{align}
This identity follows from noting that the complex part of the spherical harmonic comes solely from $\exp [ i m\phi]$ and flipping $m$ is equivalent to conjugation. However then the associated Legendre polynomial part of the spherical harmonic changes from $P_{\ell}^{m}$ to $P_{\ell}^{-m}$ under a flip of $m$, necessitating the compensating phase factor $(-1)^m$ above.

Under parity $\hat{P}$, spherical harmonics behave as
\begin{align}
\hat{P} Y_{\ell m}(\hat{r}) =  Y_{\ell m}(-\hat{r}) = (-1)^{\ell} Y_{\ell m}(\hat{r}).
\label{eqn:sph_harm_parity}
\end{align}

The plane wave expansion writes a complex exponential as an infinite sum over products of spherical harmonics and spherical Bessel functions (e.g. \citealt{AWH13}):
\begin{align}
e^{i\vk \cdot \vr} = \sum_{L} i^L j_L(kr) \sum_{M = -L}^{L} Y_{LM}(\hk)Y_{LM}^*(\hr).
\label{eqn:plane_wave_exp}
\end{align}
It can be understood as a convenient ``Rosetta stone'' enabling translation between problems with translation invariance (homogeneity) and rotation invariance (isotropy). Cosmological settings frequently involve both, rendering this identity highly useful.

The Gaunt integral is defined
\begin{align}
&\mathcal{G}_{l_1 l_2 l_3}^{m_1 m_2 m_3} \equiv \int d\Omega \;Y_{l_1 m_1}(\hat{r}) Y_{l_2 m_2}(\hat{r}) Y_{l_3 m_3}(\hat{r})  \nonumber\\
&=\mathcal{C}_{l_1 l_2 l_3} \left(\begin{array}{ccc}
l_1 & l_2 & l_3\\
0 & 0 & 0
\end{array}\right)
\left(\begin{array}{ccc}
l_1 & l_2 & l_3\\
m_1 & m_2 & m_3
\end{array}\right)
\label{eqn:Gaunt_defn}
\end{align}
with \
\begin{align}
\mathcal{C}_{l_1 l_2 l_3} \equiv \sqrt{\frac{(2l_1 + 1)(2l_2 + 1)(2l_3 + 1)}{4\pi}};
\label{eqn:C3_defn}
\end{align}
the $2 \times 3$ matrices are Wigner 3j-symbols (e.g. \citealt{NIST_DLMF} \S 34.2, with basic properties in \S34.3; more detailed discussion is in \cite{VMK}).

The integral of four spherical harmonics can be obtained by first linearizing two spherical harmonics into a sum over single spherical harmonics using the Gaunt integral. We begin with 
\begin{align}
Y_{l_1 m_1}(\hat{r}) Y_{l_2 m_2}(\hat{r}) = \sum_{LM} c_{LM}(l_1, l_2;m_1, m_2) Y_{LM}(\hat{r}),
\label{eqn:ylm_series}
\end{align}
where the coefficients $c_{LM}(l_1, l_2;m_1, m_2)$ are given by integrating both sides against $Y^*_{LM}(\hat{r})$ and invoking orthogonality, so that
\begin{align}
c_{LM}(l_1, l_2;m_1, m_2) = (-1)^M\mathcal{G}_{l_1 l_2 L}^{m_1 m_2 -M}.
\label{eqn:cLM}
\end{align}
We then have
\begin{align}
&\mathcal{H}_{l_1 l_2 l_3 l_4}^{m_1 m_2 m_3 m_4} \equiv \int d\Omega\; Y_{l_1 m1}(\hat{r}) Y_{l_2 m_2}(\hat{r}) Y_{l_3 m_3}(\hat{r}) Y_{l_4 m_4}(\hat{r})\nonumber\\
&= \sum_{L} (-1)^M\mathcal{G}_{l_1 l_2 L}^{m_1 m_2 -M} \mathcal{G}_{L l_3 l_4}^{M m_3 m_4}
\label{eqn:H_defn}
\end{align}
by inserting equation (\ref{eqn:ylm_series}) with the definition (\ref{eqn:cLM}) into the first line above and then integrating. We note that there is no sum over $M$ because it is set by the zero-sum rule on the spins enforced by the Gaunt integrals: $m_1 + m_2 = M$. We further note that the sum over $L$ has compact support because of the triangle rules on total angular momenta: $|l_1 -l_2|\leq L\leq l_1 + l_2$ and the same constraint holds replacing  $l_1\to l_3$ and $l_2\to l_4$. Finally, we chose to couple the first two angular momenta to $L$ and the last two to $L$, but of course choosing other combinations to recouple would lead to the same result.

The Gegenbauer polynomials can be written explicitly as 
\begin{align}
&C_{\ell}^{(\lambda)}(x) =  \nonumber\\
&\left(\begin{array}{c}
\ell + 2\lambda -1 \\
\ell 
\end{array}\right) \sum_{k = 0}^{\ell} \left(\begin{array}{c}
\ell \\
k 
\end{array}\right) \frac{(2\lambda + \ell)^{(k)}}{(\lambda + 1/2)^{(k)}} \frac{1}{2^k} \left(\frac{x-1}{2}\right)^k,
\label{eqn:Gbauer_exp}
\end{align}
e.g. \cite{Kim:2012} or \cite{Durand:1976}. The rising factorial is defined as $a^{(k)} = a(a+1)(a+2)\cdots (a + k -1)$.\footnote{\url{http://mathworld.wolfram.com/RisingFactorial.html}.}

\section{Factorization of Gegenbauer polynomials into spherical harmonics}
\label{app:g_add_thm}
Here we show that the Gegenbauer polynomials can be written as a sum of products of spherical harmonics each with one unit vector as their argument. The Gegenbauer polynomials have an explicit, finite representation in powers of their argument, as
\begin{align}
&C_{\ell}^{(\lambda)}(x) = \sum_{i=0}^{\ell} t_i^{\ell, \lambda} x^i,\nonumber\\
&t_i = \alpha_{\ell} \sum_{k = 0}^{\ell} \left(\begin{array}{c}
\ell \\
k 
\end{array}\right) \frac{(2\lambda + \ell)^{(k)}}{(\lambda + 1/2)^{(k)}} \frac{1}{2^k}   \left(\begin{array}{c}
k\\
i  
\end{array}\right)(-1)^{k-i},\nonumber\\
&\alpha_{\ell} = \left(\begin{array}{c}
\ell + 2\lambda -1 \\
\ell 
\end{array}\right)
\label{eqn:Gbauer_to_power}
\end{align}
where the superscripts $\ell,\lambda$ identify the Gegenbauer polynomial being expanded. The second line comes from expanding $[(x-1)/2]^k$ in equation (\ref{eqn:Gbauer_exp}) using the binomial theorem and comparing with the first line above. 

Powers can be represented using Legendre polynomials as
\begin{align}
&x^{n}  =\sum_{\ell = n, n-2, \cdots} s^n_{\ell} \mathcal{L}_{\ell}(x),\nonumber\\
&s^n_{\ell} = \frac{(2\ell + 1)n!}{2^{(n-\ell)/2} [ (1/2)(n - \ell)]! (\ell + n + 1)!!},
\label{eqn:power_to_legend}
\end{align}
where superscript $n$ identifies the power being expanded.\footnote{\url{http://mathworld.wolfram.com/LegendrePolynomial.html}, equation (15), originally from Schmied (2005) personal communication to E. Weisstein.} The double factorial has a different definition for even versus odd arguments; see e.g. equation (1) of Weisstein (2018).\footnote{\url{http://mathworld.wolfram.com/DoubleFactorial.html}}

Inserting equation (\ref{eqn:power_to_legend}) into equation (\ref{eqn:Gbauer_to_power}) we find
\begin{align}
&C_{\ell}^{(\lambda)}(x) = \sum_{j=0}^{\ell}  v_j^{\ell, \lambda} \mathcal{L}_{j}(x),\nonumber\\
&v_j^{\ell, \lambda} = \sum_{i = 0}^{\ell} t_i^{\ell} s_{j}^i.
\end{align}
Summarizing, the Gegenbauer polynomials can be written as a finite series of powers and each power written as a finite series of Legendre polynomials, leading to a double sum. The Legendre expansion coefficients of a given Gegenbauer polynomial can then be obtained by holding the order in the second series fixed, so that we take a sum over all powers that contribute to a particular Legendre polynomial as it enters the Gegenbauer expansion. This approach can be conveniently understood as writing down a matrix showing the coupling between powers and Legendre polynomials in the series for the Gegenbauer polynomial, with the row being the power and the column being the Legendre. The way we constructed the representation was by staying in a fixed row (power) and reading across columns to obtain the Legendre coefficients for that power, but to obtain the desired Legendre coefficients we now instead read across rows at fixed column.

Our final step is to apply the spherical harmonic addition theorem (\ref{eqn:sph_add_thm}) to write the Gegenbauer as
\begin{align}
C_{\ell}^{(\lambda)}(x) = \sum_j^{\ell} v_j^{\ell, \lambda}\frac{4\pi}{2j + 1}\sum_{s = -j}^j Y_{js}(\hat{a})Y_{js}^*(\hat{b})
\end{align}
if $x\equiv \hat{a}\cdot \hat{b}$. For more compact notation in the main text, we define $w_j^{\ell, \lambda} = 4\pi v_j^{\ell, \lambda}/(2j + 1)$, so that
\begin{align}
C_{\ell}^{(\lambda)}(x) = \sum_j^{\ell} w_j^{\ell, \lambda}   \sum_{s = -j}^j Y_{js}(\hat{a})Y_{js}^*(\hat{b}),
\label{eqn:C_thm}
\end{align}
though we emphasize that $w_j^{\ell, \lambda}$ is spin-independent.

\section{Proof of decoupling integral for $|\vec{p}_1 + \vec{p}_2|^{-1}$}
\label{app:dec_int_1_proof}
Here we show that the radial part of the Legendre series can be represented in terms of the overlap integral of two spherical Bessel functions, i.e. 
\begin{align}
\frac{2}{\pi}\int dx\;j_{\ell}(ax)j_{\ell}(ax) = \frac{b^{\ell}}{a^{\ell + 1}},\;\;b<a;\;\;\frac{a^{\ell}}{b^{\ell + 1}},\;\; a<b.
\end{align}
We begin with the identity 
\begin{align}
&j_L(k|\vr_1-\vr_2|)Y_{LM}(\reallywidehat{\vr_1-\vr_2})=4\pi\sum_{L_1M_1}\sum_{L_2M_2}i^{L_2-L_1+L}\nonumber\\
&\times j_{L_1}(kr_1)j_{L_2}(kr_2)\mathcal{C}_{L_1 L_2 L}\left(\begin{array}{ccc}
L_{1} & L_{2} & L\\
0 & 0 & 0
\end{array}\right)\nonumber\\
&\times \left(\begin{array}{ccc}
L_{1} & L_{2} & L\\
M_1 & M_2 & M
\end{array}\right) Y_{L_1 M_1}^*(\hr_1)Y_{L_2 M_2}^*(\hr_2),
\label{eqn:sep_id}
\end{align}
This identity is equation (46) of \cite{Slepian_RSD_model:2016} and was proven there using two different ways to write the plane wave, expanding each side into spherical harmonics and spherical Bessel functions with the plane wave expansion (\ref{eqn:plane_wave_exp}), and integrating over solid angle. The $L=0$ case of this identity is the version of Gegenbauer's addition theorem applying to spherical Bessel functions (see \citealt{Sack:1964} equation 59). $C_{L_1 L_2 L}$ is defined in equation (\ref{eqn:C3_defn}); the $2\times 3$ matrices are Wigner 3j-symbols.

Setting $L = 0$ means that the Wigner 3j-symbols force $L_1 = L_2$, and $L=0$ means $M = 0$ so that $M_2 = -M_1$. We find
\begin{align}
&j_0(k|\vr_1-\vr_2|)\frac{1}{\sqrt{4\pi}}=4\pi\sum_{L_1M_1} j_{L_1}(kr_1) j_{L_1}(kr_2)  \mathcal{C}_{L_1 L_1 0} \nonumber\\
&\times (2L_1 + 1)^{-1} Y_{L_1 M_1}^*(\hr_1)Y_{L_1 M_1}(\hr_2),
\end{align} 
where we used that $Y_{00} = 1/\sqrt{4\pi}$ and evaluated the zero-spin 3j-symbol as $1/\sqrt{2L_1 + 1}$ and the 3j-symbol with spins (with $M_2 = -M_1$) as $(-1)^{M_1}/\sqrt{2L_1 + 1}$ using \cite{NIST_DLMF} \S34.3.1. We then used the conjugate identity to rewrite the second spherical harmonic from the previous simplification as unconjugated and cancel the $(-1)^{M_1}$ phase factor. Simplifying using that $\mathcal{C}_{L_1 L_1 0} = (2L_1 + 1)/\sqrt{4\pi}$, we obtain
\begin{align}
&j_0(k|\vr_1-\vr_2|)\frac{1}{\sqrt{4\pi}}=4\pi\sum_{L_1M_1} j_{L_1}(kr_1) j_{L_1}(kr_2)  \frac{2L_1+1}{\sqrt{4\pi}} \nonumber\\
&\times (2L_1 + 1)^{-1} Y_{L_1 M_1}^*(\hr_1)Y_{L_1 M_1}(\hr_2).
\end{align} 
Canceling the $1/\sqrt{4\pi}$ and then resumming the right-hand side using the spherical harmonic addition theorem (\ref{eqn:sph_add_thm}), we find
\begin{align}
j_0(k|\vr_1-\vr_2|) = \sum_{L_1} (2L_1 + 1) j_{L_1}(kr_1)j_{L_1}(kr_2) \mathcal{L}_{L_1}(\hr_1\cdot\hr_2).
\end{align}
We now integrate both sides with respect to $k$ as
\begin{align}
&\int dk\;j_0(k|\vr_1-\vr_2|) = \frac{\pi}{2|\vr_1-\vr_2|} \nonumber\\
&=\sum_{L_1} (2L_1 + 1)\mathcal{L}_{L_1}(\hr_1\cdot\hr_2) \int dk\;j_{L_1}(kr_1)j_{L_1}(kr_2) .
\end{align}
To obtain the first line above, we simply changed variables and used that the Sine integral has value $\pi/2$. We now observe that the middle expression above may be written as a Legendre series, so we find
\begin{align}
\frac{\pi}{2}\sum_L \frac{r_<^L}{r_>^{L+1}}\mathcal{L}_{L}(\hr_1 \cdot \hr_2) =& \sum_{L_1} (2L_1 + 1) \mathcal{L}_{L_1}(\hr_1\cdot\hr_2) \nonumber\\
&\times \int dk\; j_{L_1}(kr_1)j_{L_1}(kr_2).
\end{align}
Integrating both sides against $[(2L+1)/2]\mathcal{L}_{L}(\mu_{12})$ with $\mu_{12} \equiv  \hr_1 \cdot \hr_2$, we invoke the orthogonality of the Legendre polynomials (the pre-factor is to compensate for the fact that they are not orthonormal) to find
\begin{align}
\frac{r_<^L}{r_>^{L+1}}= \frac{2}{\pi} (2L + 1) \int dk\; j_{L_1}(kr_1)j_{L_1}(kr_2)
\label{eqn:first_int_result}
\end{align}
as desired. This result also appears in \cite{Bloomfield:2017} equation (14), \cite{Mehrem:2009} equation (4.13), and in \cite{Watson}. An alternative derivation is to use the plane-wave expansion into sBFs and Legendre polynomials from the beginning, and then use the spherical harmonic addition theorem to integrate over the Legendres that will appear on the righthand side with arguments $\hat{k} \cdot\hat{r}_1$ and $\hat{k} \cdot\hat{r}_2$.

\section{Proof of decoupling integral for $|\vec{p}_1 + \vec{p}_2|^{-2}$}
\label{dec_int_2_proof}
We now prove that
\begin{align}
&\frac{2}{\pi}\int x dx\;j_{\ell+1}(ax) j_{\ell}(bx)  = \frac{b^{\ell}}{a^{\ell + 2}},\;\;b<a;\;\;0,\;\;a<b.
\label{eqn:second_des_int}
\end{align}
We may prove this second decoupling integral using the first, proven in the previous section, and the recursion relation for spherical Bessel functions. We first assume that $a>b$. We invoke the recursion relation (\citealt{NIST_DLMF} \S10.51.2)\footnote{\url{https://dlmf.nist.gov/10.51}}
\begin{align}
j_{n}'(u) = -j_{n+1}(u) +\frac{n}{u}j_n(u),
\end{align}
where prime denotes $d/du$, to find that
\begin{align}
j_{n+1}(u)  =  \frac{n}{u}j_n(u) - j_{n}'(u).
\end{align}
We set $u = ax$, so that $d/du = (1/x)\partial/\partial a$. Replacing $j_{\ell +1}(ax)$ in our desired integral (\ref{eqn:second_des_int}) and simplifying, we find the equality
\begin{align}
&\frac{2}{\pi}\int x dx\;j_{\ell+1}(ax) j_{\ell}(bx) =\frac{2}{\pi} \left[\frac{\ell}{a}- \frac{\partial}{\partial a} \right] \int dx\;j_{\ell}(ax) j_{\ell}(bx) \nonumber\\
&=\frac{b^{\ell}}{a^{\ell + 2}},\;\; b < a,
\label{eqn:second_dec_proof}
\end{align}
where the last line followed by using our result (\ref{eqn:first_int_result}) from the previous section with $a>b$ (which we assumed above) and simplifying.

We now address the case $a < b$. The first equality of equation (\ref{eqn:second_dec_proof}) still holds, but we now use the $a < b$ result from the previous section to replace the matched-order integral, finding
\begin{align}
\frac{2}{\pi} \left[\frac{\ell}{a}- \frac{\partial}{\partial a} \right] \int dx\; j_{\ell}(ax) j_{\ell}(bx) = \frac{1}{2\ell + 1}\frac{a^{\ell - 1}}{b^{\ell + 1}}\left[\ell - \ell \right] = 0.
\end{align}
We note that our eigenfunction expansion, based on this integral, offers a factorized representation of a Heaviside step function centered at $a$. We have
\begin{align}
&f(a, b) =\frac{2b^2}{\pi a} \int_0^{\infty} x\; x j_0(ax) j_1(bx) = 0,\;\;b<a;\;\;=1,\;\;b>a\nonumber\\
&\simeq H(b-a)\nonumber\\
&=\frac{2b^2}{\pi a}\sum_{n=0}^{\infty} \epsilon_n\phi^{2-}_{n0}(a) \phi^{2+}_{n1}(b). 
\end{align}
The above (save for the value at $b=a$, which is not $1/2$) is a Heaviside function $H$ shifted so that it switches on at $b = a$ rather than zero. We used the notation $\simeq$ to indicate the difference from the conventional definition of $H$ at $a=b$.

\section{Integral-to-sum identity for $|\vec{p}_1 + \vec{p}_2|^{-2}$ decoupling integral}
\label{app:identity_relax}
We now convert the spherical Bessel functions to Bessel functions (the relation is given below equation \ref{eqn:dec_int_1}). We then use \cite{Dominici:2012} Corollary 3.2, which is
\begin{align}
\int_0^{\infty} \frac{J_{\mu}(at) J_{\nu}(bt)}{t^{\mu + \nu - 2k}} dt  = \sum_{n = 0}^{\infty} \epsilon_n \frac{J_{\mu}(an) J_{\nu}(bn)}{n^{\mu + \nu - 2k}},
\label{eqn:cor_2}
\end{align}
with $0<b<a<\pi$, ${\rm Re}(\mu + \nu - 2k) > -1$ and $k$ a natural number.  We need this identity to correctly reproduce the two integrals in our expression (\ref{eqn:second_dec_int}) both for $p_2 < p_1$ {\it and} $p_1 < p_2$ however. In particular we want the vanishing of one of the two integrals at all times to be correctly reproduced by the sums. However, the corollary above does have sufficient freedom to do this, because there is no constraint between the magnitudes of sBFs' orders. 

Consider our first integral in (\ref{eqn:second_dec_int}), which turns on when $p_1 > p_2$ and is zero otherwise. We consider the two cases and show that the sum identity covers both. First, suppose $p_1 > p_2$, so we set $a = p_1$, $b=p_2$, and $\mu = \ell + 3/2$, $\nu = \ell + 1/2$. $\mu + \nu  = 2\ell + 2$ so to get the right power in the denominator (zero, as the factor of $x$ in the integrand is canceled when we convert to Bessel functions, each of which brings $1/\sqrt{x}$), we set $k = \ell + 1$.

We then have 
\begin{align}
&\int_0^{\infty} J_{\ell + 3/2}(p_1 t) J_{\ell + 1/2}(p_2 t) = \nonumber\\
&\sum_{n= 0}^{\infty} \epsilon_n J_{\ell + 3/2}(p_1 n) J_{\ell + 1/2}(p_2 n),\;\;p_1 > p_2.
\end{align}

We now consider the case where $p_1 < p_2$ (the integral vanishes here). In equation (\ref{eqn:cor_2}) we now set $a = p_2$ and $b = p_1$; but we can also set $\nu = \ell + 3/2$ and $\mu = \ell + 1/2$. We then have 
\begin{align}
&\int_0^{\infty} J_{\ell + 3/2}(p_1 t) J_{\ell + 1/2}(p_2 t) = \nonumber\\
&\sum_{n= 0}^{\infty} \epsilon_n J_{\ell + 3/2}(p_1 n) J_{\ell + 1/2}(p_2 n),\;\;p_1 < p_2.
\end{align}
The same approach shows that the second integral in equation (\ref{eqn:second_dec_int}) is also faithfully represented by the sum on both regions; the proof is exactly as above but replacing everywhere $p_1$ with $p_2$ and $p_2$ with $p_1$, so that $p_2$ is paired with the larger-index Bessel function. 

Finally, we note that \cite{Dominici:2012} give an alternative proof of Corollary 3.2 in \S4 of their paper (suggested to them by T. Koornwinder) that uses Parseval's relation for the Fourier transform and for Fourier series to show that the analog of the corollary where the integral runs from $-\infty$ to $\infty$ and the sum from $-\infty$ to $\infty$ as well holds for $0 < a, b < \pi$, i.e. with no restriction on $b$ relative to $a$ (see \citealt{Dominici:2012} equation 4.2). Since our particular case has an even integrand (the sum of the Bessel function indices is even) and an even summand, one can divide both sides by two. Properly treating the case where $n=0$ using the definition of $\epsilon_n$ as $1/2$ at $n=0$ in \cite{Dominici:2012}, this argument shows that the $a<b$ restriction is not relevant for our case.

\section{Inverse Fourier Transform of $1/|\vec{q}|^n,\;N=2,4$ and regularization}
\label{app:simple_kernel_ft}
Here we obtain the two inverse FTs required to perform the radial (over momentum magnitudes) convolution integrals as products in configuration space. See for instance equation (\ref{eqn:gn3}). The $1/q^2$ case is simple; the $1/q^4$ case is formally divergent, but the multiplication by other functions, in particular the correlation function, in the subsequent configuration space radial integrals will effectively regularize this divergence. Thus we will compute the inverse transform of a softened version of this kernel and take the softening to zero after the configuration-space radial integrals have been performed. 

For the first kernel, we have
\begin{align}
K^{[2]}(r) \equiv {\rm FT}^{-1}\left\{\frac{1}{q^2} \right\} (r) = \int \frac{q^2 dq}{2\pi^2}\;j_0(qr)\frac{1}{q^2} = \frac{1}{4\pi r}, 
\label{eqn:simple_kernel_1}
\end{align}
where the 3D inverse FT reduced to a 1D $j_0$ transform due to the spherical symmetry of the kernel. We used the change of variables $u = qr$ and then that the sine integral $\int du\;\sin u/u = \pi/2$. 

For the second kernel, we have
\begin{align}
K^{[4]}(r;\epsilon) \equiv {\rm FT}^{-1}\left\{\frac{1}{(q + \epsilon)^4} \right\} (r) = \int \frac{q^2 dq}{2\pi^2}\;j_0(qr)\frac{1}{(q +\epsilon)^4}.
\label{eqn:simple_kernel_2}
\end{align}
We note that we chose to regularize the divergence by softening the kernel here; we could also have chosen to cut off the integral above zero. We present the result of that as well, on the view that one or the other may prove more suitable for numerical work.

To evaluate the integral (\ref{eqn:simple_kernel_2}), we notice that $(1/2)\partial^2/\partial \epsilon^2$ of our previous integral (\ref{eqn:simple_kernel_1}) will generate the integral (\ref{eqn:simple_kernel_2}). To perform this shifted-denominator version of our previous integral, we change variables so that $u = q+ \epsilon$. We then in the numerator have $\sin [(u-\epsilon)r] (u-\epsilon)^2$ and can expand this using respectively the sum-to-product identity for trigonometric functions and the binomial theorem. 

We find
\begin{align}
&\int \frac{q^2 dq}{2\pi^2}\;j_0(qr)\frac{1}{( q+ \epsilon)^2}=\frac{1}{4\pi^2 r}\bigg[2{\rm Ci}(r\epsilon)  ( r\epsilon \cos r\epsilon + \sin r\epsilon) \nonumber\\
&+ (\cos r\epsilon - r\epsilon \sin r\epsilon) (\pi - 2 {\rm Si}(r\epsilon)) \bigg]
\label{eqn:f3}
\end{align}
where ${\rm Ci}$ is the cosine integral and ${\rm Si}$ the sine integral. Performing the parametric differentiation on equation (\ref{eqn:f3}), we have
\begin{align}
&\int \frac{q^2 dq}{2\pi^2}\;j_0(qr)\frac{1}{( q+ \epsilon)^4} = \frac{1}{2} \frac{\partial^2}{\partial \epsilon^2}\int \frac{q^2 dq}{2\pi^2}\;j_0(qr)\frac{1}{( q+ \epsilon)^2}\nonumber\\
&=\frac{1} {8\pi^2 \epsilon} \bigg\{4 + r\epsilon \sin r\epsilon \big[ -6{\rm Ci}(r\epsilon) + r\epsilon (\pi - 2 {\rm Si}(r\epsilon)) \big] \nonumber\\
&+ r\epsilon \cos r\epsilon \big[-3\pi - 2 r\epsilon {\rm Ci}(r\epsilon) + 6 {\rm Si} (r\epsilon) \big] \bigg\}
\end{align}

As noted above, a second regularization would have been just to truncate the integral at $\epsilon$ rather than integrate all the way down to zero:
\begin{align}
&\int_{\epsilon}^{\infty} \frac{q^2 dq}{2\pi^2}\;j_0(qr)\frac{1}{q^4}  = - \frac{r}{8\pi} + \frac{\cos r\epsilon}{4\pi^2 \epsilon} + \frac{\sin r\epsilon}{4\pi^2 r \epsilon^2} + \frac{r {\rm Si} (r\epsilon)}{4\pi^2}.
\end{align}
We obtained this result using integration by parts.



\bsp	
\label{lastpage}
\end{document}